\documentclass[sigconf]{acmart}

\usepackage{hyperref}
\usepackage[hyphenbreaks]{breakurl}
\usepackage{url}
\usepackage{color}
\usepackage{caption}
\usepackage{subfigure}
\usepackage{multirow}
\usepackage{tablefootnote}
\usepackage{makecell}
\usepackage{graphicx}
\usepackage{listings}
\usepackage[inline]{enumitem}
\usepackage{multicol}
\usepackage{xcolor}
\usepackage{enumitem}
\usepackage{xspace}
\usepackage[tableposition=below]{caption}

\newcommand{\para}[1]{{\vspace{3pt} \noindent \textbf{#1} \hspace{6pt}}}

\newcommand*\circled[1]{\kern-2.5em%
  \put(0,4){\color{blue}\circle*{18}}\put(0,4){\circle{16}}%
  \put(-3,0){\color{white}\bfseries\large#1}~~}

\newcommand{\K}{\textsc{k}\xspace}

\copyrightyear{2025}
\acmYear{2025}
\setcopyright{cc}
\setcctype{by}
\acmConference[IMC '25] {Proceedings of the 2025 ACM Internet Measurement Conference}{October 28--31, 2025}{Madison, WI, USA.}
\acmBooktitle{Proceedings of the 2025 ACM Internet Measurement Conference (IMC '25), October 28--31, 2025, Madison, WI, USA} 
\acmDOI{10.1145/3730567.3732913}
\acmISBN{979-8-4007-1860-1/2025/10}

\ccsdesc[500]{Information systems~Web crawling}

\keywords{Robots.txt; AI Crawlers; Web Content Control; Content Creators;}

\iftrue
\newcommand{\fixme}[1]{{\color{magenta} #1}}
\newcommand{\todo}[1]{{\color{red} TODO: #1}}
\newcommand{\elisa}[1]{\textcolor{blue}{\noindent[EL: #1]}}
\newcommand{\alex}[1]{\textcolor{red}{\noindent[AL: #1]}}
\newcommand{\geoff}[1]{\textcolor{cyan}{Geoff: #1}}
\newcommand{\ol}[1]{{\color{blue} #1}}

\newcommand{\eg}{{e.g.\xspace}}

\newcommand{\revision}[1]{{\color{black} #1}}

\else
\newcommand{\fixme}[1]{}
\newcommand{\todo}[1]{}
\newcommand{\alex}[1]{}
\newcommand{\elisa}[1]{}
\newcommand{\geoff}[1]{}
\newcommand{\ol}[1]{}

\newcommand{\eg}{{e.g.\ }}

\newcommand{\revision}[1]{{\color{black} #1}}

\fi

\newenvironment{packed_itemize}{
\begin{list}{\labelitemi}{\leftmargin=0.90em}
  \setlength{\itemsep}{1pt}
  \setlength{\parskip}{0pt}
  \setlength{\parsep}{0pt}
  \setlength{\headsep}{0pt}
  \setlength{\topskip}{0pt}
  \setlength{\topmargin}{0pt}
  \setlength{\topsep}{0pt}
  \setlength{\partopsep}{0pt}
}{\end{list}}

\begin{document}

\title{Somesite I Used To Crawl: Awareness, Agency and Efficacy in Protecting Content Creators From AI Crawlers}

\author{Enze Liu}
\email{e7liu@ucsd.edu}
\affiliation{%
  \institution{UC San Diego}
  \city{La Jolla}
  \state{CA}
  \country{USA}
}
\authornote{Equal contribution.}
\author{Elisa Luo}
\email{e4luo@ucsd.edu}
\affiliation{%
  \institution{UC San Diego}
  \city{La Jolla}
  \state{CA}
  \country{USA}
}
\authornotemark[1]
\author{Shawn Shan}
\email{shansixiong@cs.uchicago.edu}
\affiliation{%
  \institution{University of Chicago}
  \city{Chicago}
  \state{IL}
  \country{USA}
}
\author{Geoffrey M.\ Voelker}
\email{voelker@ucsd.edu}
\affiliation{%
  \institution{UC San Diego}
  \city{La Jolla}
  \state{CA}
  \country{USA}
}
\author{Ben Y.\ Zhao}
\email{ravenben@cs.uchicago.edu}
\affiliation{%
  \institution{University of Chicago}
  \city{Chicago}
  \state{IL}
  \country{USA}
}
\author{Stefan Savage}
\email{ssavage@ucsd.edu}
\affiliation{%
  \institution{UC San Diego}
  \city{La Jolla}
  \state{CA}
  \country{USA}
}
\renewcommand{\shortauthors}{Enze Liu et al.}

\begin{abstract}

  The success of generative AI relies heavily on training on data scraped
  through extensive crawling of the Internet, a practice that has raised
  significant copyright, privacy, and ethical concerns. While few measures
  are designed to resist a resource-rich adversary determined to scrape a
  site, crawlers can be impacted by a range of existing tools such as
  robots.txt, NoAI meta tags, and active crawler blocking by reverse proxies.

  In this work, we seek to understand the ability and efficacy of today's
  networking tools to protect content creators against AI-related
  crawling. For targeted populations like human artists, do they have the
  technical knowledge and agency to utilize crawler-blocking tools such as
  robots.txt, and can such tools be effective? Using large scale measurements
  and a targeted user study of 203 professional artists, we find strong
  demand for tools like robots.txt, but significantly constrained by
  critical hurdles in technical awareness, agency in deploying them, and limited
  efficacy against unresponsive crawlers. We further test and evaluate
  network level crawler blockers provided by reverse proxies. 
  Despite
  relatively limited deployment today, they offer stronger protections against AI crawlers, but still come with 
  their own set of limitations.

\end{abstract}

\maketitle

\section{Introduction}

The success of generative AI relies heavily on training on data scraped
through extensive crawling of the Internet, a practice that has raised
significant copyright, privacy, and ethical concerns. Today, AI model
trainers have unleashed large numbers of data crawlers on the Internet. By
many reports, these crawlers now dwarf the volume of human traffic on the
Internet, partly because human users consume content at a much lower rate
than crawlers.
\revision{For example, analysis by Akamai and Imperva suggest that roughly 50--70\% of website traffic 
is due to automated crawlers~\cite{akamaistudy,impervabadbots}}.
Other anecdotal evidence suggests that AI crawlers are effectively producing DDoS attacks on
smaller websites~\cite{aiddos,aiddos2}. 


While Internet crawling is well-studied, the widespread adoption of
generative AI and its intensive data scraping has significantly changed the
landscape. Data creators and hosting platforms, who were generally ambivalent
about having their content crawled in the past, are now raising serious concerns about
AI-related crawling, particularly regarding copyright, privacy, and ethical
practices. Indeed, these concerns have manifested in over thirty ongoing copyright
lawsuits~\cite{case-tracker,getty-sue,class-action}, multiple data strikes~\cite{artstation-protest,protest2}, and a surge in the adoption
of anti-crawling tools~\cite{longpre2024consent}. 

Given this new tension between AI training companies seeking training data
and content creators who consider unauthorized AI training an existential
threat to their livelihoods~\cite{statement-on-training},
a natural question arises: {\em What tools, if any, can content creators use
  to prevent their content from being crawled for AI training?} 
Answering this question requires a more thorough understanding of the needs
of content creators; their awareness of, accessibility to, and agency over
anti-crawling mechanisms; and ultimately, the availability and efficacy of
current tools. 

This paper presents our efforts to address these issues from several
complementary perspectives. In terms of representative content creators, we focus
on visual artists as the most vulnerable population being targeted by AI
crawlers. In terms of anti-crawling mechanisms, we focus on two tools at
different ends of the spectrum. The most prominent and popular
tool is robots.txt, a voluntary (and non-enforceable) protocol that enables
site owners to specify crawling restrictions. We also consider crawler
blocking by reverse proxies (e.g., Cloudflare), an active approach that
enforces blocking but has seen limited deployment. 

We begin with a longitudinal analysis of robots.txt files across the Web. Utilizing data from Common Crawl~\cite{cc:online}, we analyze the
inclusion of directives that specifically target AI crawlers over time.
This
effort
serves as broader context on how the arrival of AI crawlers has changed views
across the Web towards crawling. We then turn our attention to visual
artists, and perform a user study to understand their attitudes towards
AI crawlers, and their awareness of and accessibility to defensive tools like
robots.txt. We complement these results with measurements of 1100+ professional artist
websites to examine the hosting services artists use and levels of control these services provide.
Next, we use sites under our control to determine
which AI crawlers respect robots.txt. Finally, we consider active crawler
blocking techniques, and measure their deployment as
well as their efficacy across different AI crawlers.



Results from our study highlight critical hurdles that limit or prevent
the effective utilization of protective tools by individual creators, leaving
these key stakeholders in the data ecosystem vulnerable and often unable to
safeguard their work from unauthorized AI-driven use. More specifically, our
analysis produces a number of interesting findings:
\begin{packed_itemize}
\item We measured the inclusion of AI crawlers in robots.txt of large, popular
  sites, and found an initial surge followed by a slow increase. A
  small but growing number of websites also explicitly invite AI crawlers to crawl
  their content.
\item We conducted a survey with 203 professional artists, and found
  that individual artists often do not have the knowledge ($59\%$ have
  never heard about robots.txt) and technological means to include AI crawlers in
  their robots.txt. Once presented with more information, many artists
  indicated that they would like to use robots.txt to disallow AI crawling.
  At the same time, the majority of the artists do not trust that AI companies will respect it.
\item Testing on our own sites, most large AI companies currently do respect
  robots.txt. However, a number of AI-powered apps and crawlers do not
  respect it (including crawlers from ByteDance).
\item We measure the adoption and operation of active blocking
  mechanisms. While they offer stronger protection, they still suffer
  from limitations such as an incomplete list of AI crawlers blocked, and inability
  to stop AI training for Meta, Google, and Webzio.

\end{packed_itemize}

\noindent
Altogether, our work highlights the need for better mechanisms that account for the diverse range of use cases, that make mechanisms more accessible to a broader range of content creators, and that more clearly convey the implications and limitations of using them.




\section{Background and Related Work}


We start by providing a brief overview of AI-related crawling, and then discuss existing mechanisms that sites can use to prevent it.

\subsection{Data Scraping of Commercial AI}
\label{subsec:overview}

Crawlers are automated programs that visit websites and download their content. In the era of AI, companies use crawlers for a variety of purposes. At the time of writing, there exist three main types of AI-related crawlers: (1) crawlers for collecting training data (e.g., OpenAI's GPTBot), (2) crawlers for augmenting AI-backed assistants (e.g., OpenAI's ChatGPT-User), and (3) crawlers 
for facilitating AI-backed search engines (e.g., OpenAI's SearchBot).


\textit{Crawlers for collecting training data (AI data crawlers).} One significant use of crawlers is to collect data for training AI models.
Some companies have developed their own crawlers for such purposes, and others rely upon third-party crawlers (e.g., Common Crawl~\cite{cc:online}). 

\textit{Crawlers for augmenting AI-backed assistants (AI assistant
  crawlers).} The second significant use of crawlers is to enhance
AI-backed assistants with additional information by fetching Web
content in real time. For instance, ChatGPT-User is a crawler that can
visit websites to fetch additional information when a user poses a
question beyond ChatGPT training data. In such cases, the crawler
retrieves relevant content from the site and delivers it to the
user. While some companies, like OpenAI, state that website content
accessed by AI assistants is not directly used for training, it could
inadvertently contribute if the company trains models on user
interaction logs, as seen with ChatGPT~\cite{oai-train}.

\textit{Crawlers for facilitating AI-backed search engines (AI search crawlers).} A third major use of crawlers is to facilitate AI-backed search engines. For example, OpenAI-SearchBot is a crawler that indexes websites, which in turn is used by AI-backed search engines. While companies claim that the content of a website retrieved by AI search crawlers is not directly used for training, the user or owner of a website cannot enforce nor verify this claim.




\subsection{Mechanisms against Crawling}
\label{subsec:anti-crawl}

Next we discuss current mechanisms for controlling
crawling. We focus specifically and exclusively on data transfer-centric mechanisms designed to \emph{prevent} the acquisition of content for the purpose of training AI models, rather than content-centric mechanisms such as Glaze~\cite{shan2023glaze} that focus on limiting the value of the acquired data.


\para{Robots.txt.} The Robots Exclusion Protocol (RFC9309~\cite{rfc9309}) defines
robots.txt, allowing website owners to signal which URLs crawlers
should access. Originally designed to reduce server load, it is now
widely used to manage content access. As an honor-based system,
compliant crawlers follow its directives, but adherence is not
mandatory.  Note that this approach is distinct (and indeed opposite) from
browser-oriented mechanisms, such as Global Privacy Control (GPC)~\cite{gpc:online} and Global Privacy Platform (GPP)~\cite{gpp:online}, which are designed to
let browsers signal privacy preferences to websites (e.g., if they permit their user data to be sold to third-parties).\footnote{Both the GPC and GPP
  systems were built in response to affirmative consumer privacy obligations,
  such as provided in Europe's General Data Privacy Regulation (GDPR) and
  California's Consumer Privacy Act (CCPA).  As of yet the
  statutory legal landscape for protecting content creator interests
  has not had similarly crisp rules --- perhaps explaining the absence
  of standardized AI-use permission signaling.}

Figure \ref{fig:robotstxt} shows an example robots.txt file. The first two lines allow Googlebot to crawl all URLs, while the next three disallow ChatGPT-User and GPTBot from crawling any. The final lines block all other crawlers from accessing the \texttt{/secret/} directory. Robots.txt can also include sitemaps (URL lists for indexing).


In this paper, we categorize the levels of restriction imposed by robots.txt on a given crawler into four distinct groups. The first category, \textbf{no robots.txt}, applies to sites that do not have a robots.txt file. The second, \textbf{no restrictions}, refers to cases where the user agent is fully allowed to access the website as specified by robots.txt. The third category, \textbf{partially disallowed}, indicates that the user agent is permitted to access some paths but not all. Finally, \textbf{fully disallowed} describes instances where the user agent is prohibited from accessing any paths on the website.


\begin{figure}[t]
  \centering
  \begin{verbatim}
# An example robots.txt file
User-agent: Googlebot
Allow: /

User-agent: ChatGPT-User
User-agent: GPTBot
Disallow: /

User-agent: *
Disallow: /secret/
  \end{verbatim}
   \vspace*{-0.15in}
   \caption{
     In this example robots.txt file, Googlebot is allowed to crawl all URLs on the website, ChatGPT-User and GPTBot are disallowed from crawling any URLs, and all other crawlers are disallowed from crawling URLs under the \texttt{/secret/} directory.}
  \label{fig:robotstxt}
  \vspace{-0.15in}
\end{figure}

More recently, companies have provided managed services for robots.txt. These managers simplify maintenance by offering automated updates and interfaces. Dark Visitors~\cite{darkvisitorsmain} syncs with an AI crawler database, while tools like YoastSEO~\cite{YoastSEO:online} and AIOSEO~\cite{AIOSEO:online} provide more intuitive features for configuring rules. 


\para{Active blocking.} Active blocking prevents crawlers from
accessing a website using various methods for detecting and reacting
to crawlers.  Detection methods range from simple IP address or
user agent rules to more sophisticated techniques like browser
fingerprinting.  Once detected, a website can block the crawler by
returning an error HTTP status (e.g., 403 Forbidden),
displaying an alternative page (e.g., a CAPTCHA), or even serving fake content (e.g., Cloudflare's Labyrinth~\cite{CloudflareLabyrinth:online}).  Active blocking can
be implemented directly on a web server (e.g., via Apache or Nginx
rules) or through third-party services like Cloudflare's reverse
proxy.




\para{NoAI meta tag.} First proposed by DeviantArt, NoAI and NoImageAI
are meta tags~\cite{noAIMetaTag} a site can insert into HTML content
to indicate to crawlers that content should \textit{not} be used
for AI training:\\[0.05in]
\verb|   <meta name="robots" content="noai, noimageai">|\\[0.05in]
Previous work~\cite{dinzinger2024longitudinal} found that the
adoption of these tags is low.  We confirm this result by checking the top 10\K
domains in the Tranco ranking from October 2024, with only 17 sites having
\verb|noai| and 16 having \verb|noimageai| tags.





\para{ai.txt.} Introduced by Spawning AI, ai.txt allows content
owners to specify whether AI crawlers can use their data for
training~\cite{aitxt}. Unlike robots.txt, ai.txt is read when an AI
model attempts to download media, enabling real-time updates to
preferences, even for previously collected data. Its creators argue it
offers a legally enforceable standard, referencing the EU TDM Article 4
exception~\cite{aiTDMexception}, though its enforcement differences
from robots.txt remain unclear.

\subsection{Related Work}
Given the broad scope of our work, we survey a variety of related work in the areas of Web content control mechanisms, crawler detection and blocking, and the impact of generative AI on content creators.

\para{Web content control mechanisms.} Robots.txt, arguably the most widely-used web content control mechanism, has been extensively studied. Sun et al.~\cite{sun2007large} performed a large-scale analysis, identifying errors and the increased use of the now-deprecated ``Crawl-Delay'' field. Studies by Sun et al.~\cite{sun2007determining} and Kolay et al.~\cite{kolay2008larger} revealed biases favoring major search engines. Non-technical aspects, such as legal implications of violating robots.txt~\cite{schellekens2013robot} and its use for expressing copyright authorization~\cite{yang2010using}, have also been explored. Similar protocols, like security.txt~\cite{poteat2021you} and ads.txt~\cite{bashir2019longitudinal}, have been examined for purposes beyond Web content control.

More recently, studies have revisited robots.txt in the context of
generative AI. Dinzinger and Granitzer surveyed web content control
mechanisms~\cite{dinzinger2024survey}, and empirical
studies~\cite{dinzinger2024longitudinal,longpre2024consent} found a
sharp increase in robots.txt adoption post-generative AI, with other
mechanisms like the \verb|noai| meta tag remaining
rare. Fletcher~\cite{fletcher2024many} recently conducted a case study
on the adoption of robots.txt by news websites. 
Several blog posts have examined the use of robots.txt at small scales (e.g., hundreds of websites)~\cite{palewi:online, pressgazette:online, originalityblogpost:online, newsguardtech:online}.
These studies focus
on broad trends, while our work mainly examines the perspective of individual
creators and the unique challenges they face.

\para{Detection and blocking of Web crawlers.} Research on Web crawler detection and blocking has explored various techniques, including web traffic analysis~\cite{jacob2012pubcrawl, lourencco2006catching, iliou2019towards}, server access logs~\cite{stevanovic2012feature, iliou2021detection, rovetta2019bot}, user behavior~\cite{chu2018bot, iliou2021detection}, pattern matching~\cite{kwon2012web}, machine learning~\cite{suchacka2021efficient, jan2020throwing}, and browser fingerprinting~\cite{vastel2020fp, jonker2019fingerprint, amin2020web}. Studies have also differentiated crawler behaviors, such as good versus bad bots~\cite{li2021good}, bogus bots~\cite{bai2014analysis}, and human versus bot access patterns~\cite{alnoamany2013access, lee2009classification}. Websites use blocking methods like 403 errors, CAPTCHAs, or altered pages~\cite{pham2016understanding, amin2020web}. Our work builds on analyses of website and anti-bot service behavior, including studies by Pham et al.~\cite{pham2016understanding} on user agents, Azad et al.~\cite{amin2020web} on anti-bot service effectiveness, and Jones et al.~\cite{jones2014automated} on automated detection of block pages. 


\para{Impact of generative AI on content creators.} A third area of research investigates the impact of generative AI on content creators. The work closest to ours focuses on the impact of generative AI on artists and art. For example, the blog posts by Ortiz~\cite{Ortiz:online} and Zhou~\cite{Zhoupost:online} highlighted two specific harms created by AI art: plagiarism and loss of jobs. Jiang et al.~\cite{jiang2023ai} comprehensively categorize different types of issues raised by generative AI. More empirically, Kawakami et al.~\cite{kawakami2024impact}, Shi et al.~\cite{shi2023understanding}, Lovato et al.~\cite{lovato2024foregrounding}, Ali and Breazeal~\cite{ali2023studying}, and various reports~\cite{DACSReport, bookanartistSurvey:online} have identified similar kinds of concerns by summarizing online discussions or surveying artists. Huang et al.~\cite{huang2023generative} conducted a field experiment and found that the adoption of generative AI could adversely impact the activities of artists on digital art platforms. Zhou and Lee~\cite{zhou2024generative} measured the amount and impact of AI-assisted art activities. Shan et al.~\cite{shan2023glaze} highlighted the specific concern of style mimicry (using AI to generate a specific style of art). Lastly, others have discussed the benefits and harms of generative AI art~\cite{epstein2023art, cetinic2022understanding,park2024work,newton2023ai,garcia2024paradox} as well as studied the attitudes and sentiment toward generative AI art~\cite{bellaiche2023humans,hong2019artificial,latikka2023ai,mikalonyte2022can,johnston2024understanding,raman2024exploring}. Our work contributes to this strand of research by examining the technical needs and challenges artists face in protecting their online presence.





\begin{table*}[t]
\centering
\begin{tabular}{lllccc}

\toprule
\textbf{User Agent} & \textbf{Category} & \textbf{Company} & \textbf{Publish IP} & \textbf{Claim Respect} & \textbf{Respect in Practice} \\ 
\midrule
Amazonbot & AI Search & Amazon & Yes & Yes  & Yes\\ 
AI2Bot & AI Data & Ai2 & No & - & - \\ 
anthropic-ai & Undocumented AI & Anthropic & No & - & -\\ 
Applebot & AI Search & Apple & Yes & Yes  & Yes\\ 
Applebot-Extended* & AI Data & Apple & - & Yes  & -\\ 
Bytespider & AI Data & ByteDance & No & - & No\\ 
CCBot & AI Data & Common Crawl & Yes & Yes  & Yes\\ 
ChatGPT-User & AI Assistant & OpenAI & Yes & Yes  & Yes\\ 
Claude-Web & Undocumented AI & Anthropic & No & - & -\\ 
ClaudeBot & AI Data & Anthropic & No & Yes  & Yes\\ 
cohere-ai & Undocumented AI & Cohere & No & - & -\\ 
Diffbot & AI Data & Diffbot & No & - & - \\ 
FacebookBot & AI Data & Meta & Yes & Yes  & - \\ 
Google-Extended* & AI Data & Google & - & Yes  & -\\ 
GPTBot & AI Data & OpenAI & Yes & Yes  & Yes \\ 
Kangaroo Bot & AI Data & Kangaroo LLM & No & Yes  & -\\ 
Meta-ExternalAgent & AI Data & Meta & Yes & - & Yes\\ 
Meta-ExternalFetcher & AI Assistant & Meta & Yes & No  & -\\ 
OAI-SearchBot & AI Search & OpenAI & Yes & Yes  & -\\ 
omgili & AI Data & Webz.io & No & Yes  & -\\ 
PerplexityBot & AI Search & Perplexity & No & Yes  & -\\ 
Timpibot & AI Data & Timpi & No & - & -\\ 
Webzio-Extended* & AI Data & Webz.io & - & Yes  & -\\ 
YouBot & AI Search & You.com & No & - & -\\
\bottomrule
\end{tabular}
\caption{Summary of AI user agents studied and the companies associated with them.  We derive the category from the Dark Visitors list~\cite{darkvisitorsagents} and note whether companies publish the IP addresses they use when crawling with a particular user agent, whether their documentation claims to respect robots.txt, and whether they respect robots.txt in practice (Section~\ref{sec:robotsbehavior}). If we cannot find documentation associated with a user agent or the documentation does not mention whether they respect robots.txt, we mark it as `-'. If we cannot test whether a user agent respects robots.txt (because the crawler did not visit our website), we mark it as `-'. \hspace*{0.03in} *These three user agents are not used by real crawlers, but instead are special user agents site owners can use to control crawler behavior (Section~\ref{subsec:activeBlockGeneral}). As a result, we mark their IP address as `-'.}
\label{tab:ai_agents}
\end{table*}

Also related, but orthogonal to our work, is the study of the impact of generative AI on other communities, such as user experience design professionals~\cite{li2024user}, early-career game developers~\cite{boucher2024resistance}, comedians~\cite{mirowski2024robot}, Jewish Americans~\cite{precel2024canary}, professional playwrights~\cite{grigis2024playwriting}, 
creative writers~\cite{ilonka2024creative, guo2024pen}, and online communities such as Stackoverflow and Reddit~\cite{burtch2023consequences}.






\section{How Well-resourced Websites Reacted}
\label{sec:robotstxt}

To provide a broader context on how the arrival of AI crawlers
changed views across the Web towards crawlers, we start by revisiting
how well-resourced websites reacted. These websites are more likely to
react swiftly, as they have substantial content to protect and the
technical capability and domain knowledge to do so.

In this section, using a corpus of popular domains, we investigate the
extent to which well-resourced websites adopt robots.txt to restrict
AI-related crawlers. Among these popular sites, many are quick to add
restrictions to AI crawlers in robots.txt: over 10\% of the domains
explicitly disallowed AI crawlers in their robots.txt file after AI
crawler user agents were announced.  While there have been many
different incentives and efforts (e.g., the recent EU AI Act) to use
robots.txt to restrict AI crawlers, we also observe a small yet
noticeable reverse trend: some sites recently removed restrictions on
AI crawlers, likely due to reasons such as entering into data
licensing agreements with AI companies.

\subsection{Data and Methodology}
\label{sec:major-dataset-method}

To explore historic trends in the use of robots.txt to control AI
crawlers, we compile a comprehensive list of user agents for AI
crawlers and a longitudinal dataset of robots.txt files for sites
that are consistently popular over time.

\para{AI user agents.}
We compile a comprehensive list of AI user agents based on Dark
Visitors, an industry blog that maintains an up-to-date list of AI user
agents~\cite{darkvisitorsagents}. Since Dark Visitors also lists other
crawler user agents, we only consider the AI-related user agents belonging to
the following categories: \textsc{AI Assistant} (AI Assistant Crawler
in this paper), \textsc{AI Data Scraper} (AI Data Crawler in this
paper), \textsc{AI Search Crawler}, and \textsc{Undocumented AI
  Agents}. We also cross-validated the list with a prior study that
collected popular user agents in robots.txt
files~\cite{longpre2024consent} and confirmed that our list is a
superset of the AI user agents in this prior study. In total, we use
24 unique AI-related user agents, listed in
Table~\ref{tab:ai_agents}. We focus exclusively on these user agents
for the rest of the paper unless otherwise noted.



\para{Historic robots.txt data from Common Crawl.}
We compile a list of sites that are consistently popular over time to
represent a stable set of well-resourced websites that have
substantial valuable content, and the knowledge and resources to
control AI crawler access to it.  In particular, we focus on popular
sites whose domains appear in the Tranco Top 100\K lists every month
for two years, from October~2022 through October~2024. \revision{We
  restrict the list to sites that appear in all of the top 100\K lists
  over this period to avoid having our results affected by list
  churn~\cite{scheitle2018alongway}.}  There are 51,605 sites whose
domains consistently appear in the top 100\K lists over these two years.

For each of these sites, we look for historic robots.txt files served
by the sites in Common Crawl~\cite{cc:online} snapshots covering the October 2022--2024
period.
All snapshots crawled each site at least once; if a snapshot crawled a
site more than once, we use the most-recent robots.txt in that
snapshot.  Table~\ref{tab:CC-data} in
Appendix~\ref{appendix:cc-data-summary} lists each Common Crawl
snapshot, the months it covers, and the number of sites with a
robots.txt file.

We exclude sites that did not have robots.txt files, as well as sites
where Common Crawl encountered an error when requesting robots.txt
from them.\footnote{For instance, if a site implemented active
  blocking on automated requests (like those of the CC crawler), then
  Common Crawl may record a 403 Forbidden HTTP status code for those
  sites.}
Of the 51,605 longitudinally popular sites, 40,455 of them have
a robots.txt file in every snapshot of the Common Crawl data.  We
refer to these 40,455 sites as the Stable Top 100\K, and these are
the sites we use in our analyses.
Each Stable Top 100\K site appears in all top 100\K rankings over time
and has a robots.txt file in every Common Crawl snapshot.




We validated that the Common Crawl data is accurate by manually comparing robots.txt files retrieved by Common Crawl with the temporally closest
version available in the Internet Archive \revision{for a random
  sample of ten robots.txt files in each Common Crawl
  snapshot}. We also validated the last snapshot of the Common Crawl
data by conducting our own crawl of robots.txt of the top 10\K sites
of the Stable Top 100\K. \revision{There was no disagreement between
  the robots.txt files collected by Common Crawl and Internet Archive.
  We found minimal (<1\%) disagreements between our own crawl and
  Common Crawl, which we attribute to websites changing the contents
  of robots.txt in the time between the two crawls (the day we
  performed our crawl could be up to multiple weeks later than when
  the site appeared in the last Common Crawl snapshot).}

\para{Parsing and interpreting robots.txt.}
We parse robots.txt files using Google's robots.txt
parser~\cite{googlerobotstxtparser}. We rely on Google's parser as
robots.txt is a complex standard and our experience suggested that
home-grown parsers are error-prone.\footnote{An example is the parser
  developed by~\cite{longpre2024consent}, which we estimate to have a
  10\% error rate in parsing robots.txt. We notified the authors about
  this issue, and it has since been corrected.} We randomly selected a set of 100 robots.txt files, and
manually verified that Google's parser correctly interpreted all of
them.  We also verified that the parser correctly interpreted a
variety of edge cases not captured by other parsers, as shown in
Appendix~\ref{appendix:robotstxt-edgecase}.

We built a wrapper around Google's parser to categorize whether a
given user agent is \textit{fully disallowed} (for all content on the
site), is \textit{partially disallowed} (for a portion of the site),
or has \textit{no restrictions}.  In our analyses, we only consider a
site to disallow an AI crawler if the site's robot.txt file
has an explicit rule for the crawler's user agent.  While less than 2\%
of the domains in the Stable Top 100\K have robots.txt files with a
wildcard rule that disallows \textit{all} crawlers (e.g.,
\verb|User-agent: *|), we do not consider such sites to express an
intent to specifically disallow AI crawlers.
The code for categorizing AI user agents in robots.txt files is
publicly available at \url{https://github.com/ucsdsysnet/ai-crawler-imc-25}.

\begin{figure}[t]
  \centering
  \includegraphics[width=1.0\columnwidth]{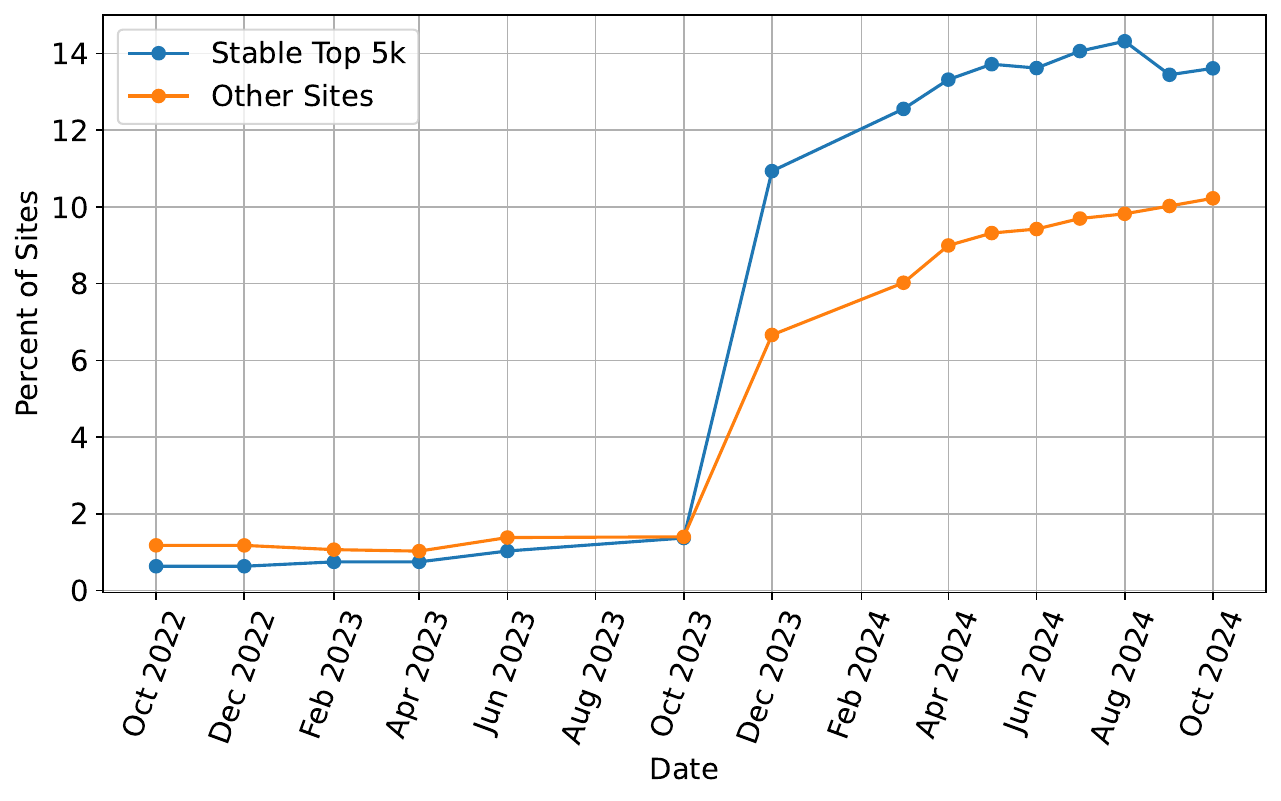}
  \caption{Percent of sites that fully disallow at least one AI crawler user agent for the Stable Top 5\K (2,551 sites) and the remaining sites in the Stable Top 100\K (37,904 sites).}
  \label{fig:by-site-rank}
    \vspace{-0.1in}
\end{figure}

\subsection{Increasing Drive to Protect Data}

Figure~\ref{fig:by-site-rank} shows the trend of restrictions on AI
crawlers over time with curves for two categories of sites: the Stable
Top 5\K sites, and all other sites in the Stable Top 100\K.  The
Stable Top 5\K sites are the 2,551 sites consistently ranked in
the top 5\K in every Tranco list throughout October~2022--2024.  While
all sites in the Stable Top 100\K have popular content and significant
resources to manage it, the Stable Top 5\K represent the very largest
sites on the Web.
Each point shows the percent of sites in a category that fully
disallow at least one AI crawler user agent in a particular Common
Crawl snapshot.  For snapshots that span multiple months, we use the
most recent month of the snapshot to represent it (e.g., points at
December 2022 correspond to the ``November/December 2022'' snapshot).

While both categories of sites have an initial surge disallowing AI
crawlers in their robots.txt after October 2023 (around the
announcement of OpenAI's GPTBot and ChatGPT-User user agents that
identify their crawlers), the most popular websites are noticeably
quicker to add restrictions in robots.txt.  Likely since they value
their content so highly, a larger proportion of the most popular sites
have restrictions on at least one AI crawler (12--14\%) when compared
to the rest of the Stable Top 100\K sites (8--10\%).
\revision{We also looked at other popularity tiers below the Stable
  Top 5\K.  In those tiers the proportions of sites that fully
  disallow AI user agents are all very similar to each other, so we
  combine them together into the ``Other Sites'' curve for clarity
  to avoid many overlapping curves.}




\begin{figure}[t]
  \centering
  \includegraphics[width=1.0\columnwidth]{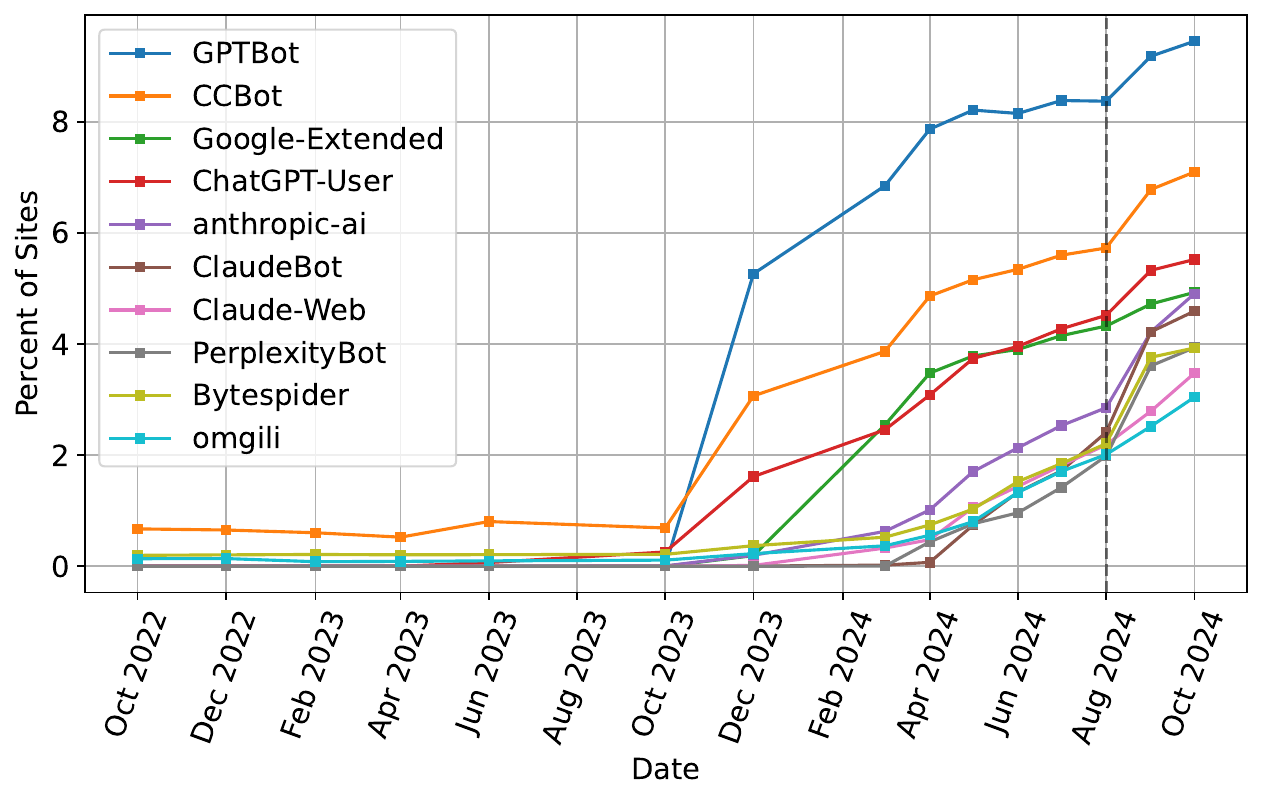}
  \caption{Percent of Stable Top 100\K sites that partially or fully
    disallow an AI crawler user agent in robots.txt over time.  The
    vertical line indicates the release of the EU AI Act.}
  \label{fig:explicit-restrictions-by-ua}
    \vspace{-0.12in}
\end{figure}



Figure~\ref{fig:explicit-restrictions-by-ua} shows historical site
robots.txt behavior for specific AI user agents.  Each curve shows the
percent of Stable Top 100\K sites that either fully or partially
disallow the corresponding AI user agent over time.
The most frequently restricted user agents are GPTBot (OpenAI) and
CCBot (Common Crawl). While Common Crawl merely collects the data (and
does not use it for any AI-related purpose itself), Common Crawl is a
very frequent data source for AI training~\cite{CCAiTrainingData}.

After August 2024 there appears a secondary distinct uptick of
restrictions for all user agents.  This uptick correlates with the
release of the EU Artificial Intelligence Act, which aims to impose
legal regulations on general-purpose AI.  Critically, the draft
version of the Act's ``Code of Practice'' explicitly requires
signatories to respect the directives of robots.txt (Sub-Measure 4.1)
to avail themselves of statutory ``Text and Data Mining'' copyright
carve-outs~\cite{AIAct-regulations}.


\subsection{Recent Decrease in Restrictions}

Among the Stable Top 5\K sites, we surprisingly not only see the trend
of adding restrictions to AI crawlers in robots.txt level off, but
also some decreases at the end of the time period. This latest
behavior is in contrast to predictions in~\cite{longpre2024consent} of
strictly increasing observable intent to disallow AI crawling.


\para{Public data licensing deals.}
One reason why a site will remove an AI crawler from their robots.txt
is when the site owner has entered into a data licensing agreement
with an AI company. A blog post from early October 2024 confirmed that
such partnerships were indeed the reason for the removal of GPTBot
from the robots.txt files from the websites of several major
publishers, including \textit{The Atlantic} and \textit{Vox
  Media}~\cite{wiredaiunblocking}. These deals often involve a
publisher who controls dozens of domains; \eg, Newscorp owns more than
10 news and media companies, each having its own set of domains.

\begin{figure}
  \centering
  \includegraphics[width=1.0\columnwidth]{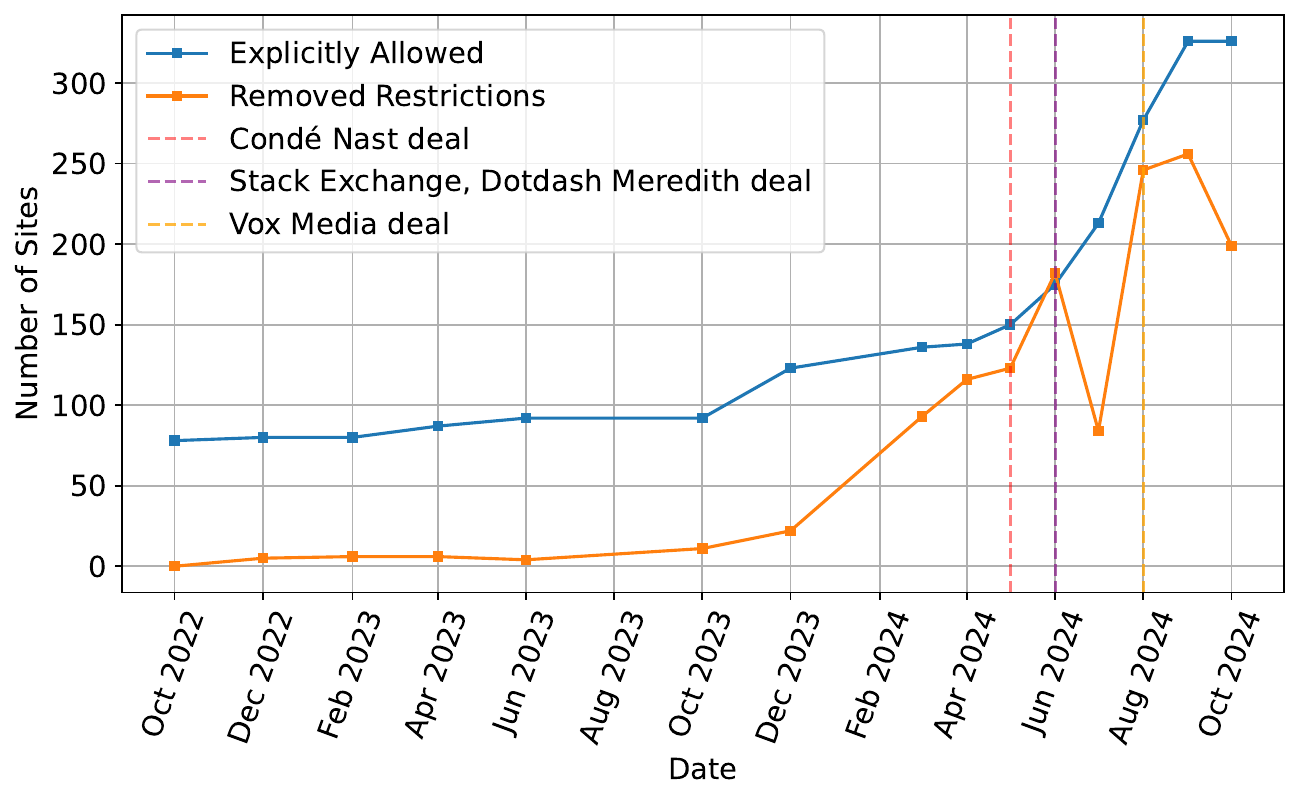}
  \caption{Number of sites that explicitly allow at least one AI crawler in their robots.txt over time, and number of sites that removed restrictions on AI crawlers in each time period. The vertical lines indicate public data deals between major publishers (who control 40+ domains) and OpenAI.}
  \label{fig:allow}
    \vspace{-0.12in}
\end{figure}

In our data, between August 2023 (the announcement of OpenAI's GPTBot
and ChatGPT-User user agents) and October 2024 (the end of our
dataset), 484 sites removed explicit restrictions on GPTBot from their
robots.txt (Figure~\ref{fig:allow}). Many of these sites are owned by
publishers who have struck publicly-announced data licensing
agreements with OpenAI, such as Dotdash
Meredith~\cite{dotdashDataDeal} (e.g., investopedia.com, people.com,
allrecipes.com), Stack Exchange~\cite{stackexchangeDataDeal} (e.g.,
superuser.com, stackoverflow.com), and Conde
Nast~\cite{condenastDataDeal} (e.g., newyorker.com, vanityfair.com,
wired.com). Some of these data usage agreements require OpenAI to
place direct links to the sites when ChatGPT generates content based
on their data, driving more traffic to their website.  The full list
of such websites is in Table~\ref{tab:detail-allow} in
Appendix~\ref{appendix:allow-robots}.

\para{Possible private deals.} In the case of major American publisher Future PLC, more than 10 of their sites (including techradar.com, tomsguide.com, and cyclingnews.com) removed restrictions on GPTBot in May 2024, while the rest of the robots.txt file remained unchanged. However, in an August 2024 podcast, the CEO of Future PLC stated that they did not have a partnership with OpenAI~\cite{futureplcPodcast}. A few other smaller publishers and news sites also removed restrictions on GPTBot, which could indicate possible private deals. 

\subsection{Recent Increase in Allowing AI Crawlers}

To our surprise, a growing number of sites explicitly allowed AI crawlers in their robots.txt, welcoming AI crawlers to scrape their content. While a small number of sites fall into this unique category, the overall number of sites that \textit{explicitly allow} AI crawlers is increasing over time as shown in Figure~\ref{fig:allow}. 


In total, 79 sites not only had no restrictions on
GPTBot in their robots.txt, but also included a rule that explicitly
allowed the GPTBot user agent.  The data licensing agreements between
OpenAI and publishers mentioned previously explain part of this increase
(especially in mid-2024), but there are also other reasons.

Among the sites that explicitly allow AI crawlers are popular
right-wing misinformation sites, which may be motivated to spread
misinformation to LLMs. Other cases are shopping sites that
potentially seek to use LLMs to increase traffic to their site.
Appendix~\ref{appendix:allow-robots} shows the full list of sites
where we observe this \textit{reverse} intent toward GPTBot. This case
study highlights that sites have a variety of motives for allowing AI
companies to crawl their data.

\section{Sentiments and Actions of Individual Artists}
\label{sec:artist-sentiment}

Section~\ref{sec:robotstxt} showed that many well-resourced websites swiftly adopted robots.txt to protect their content. In this section, we explore the question of what individual artists think about AI-related crawling and what actions they have taken in response to such crawling, if any. Compared to large organizations, individual artists have significantly more direct risk yet are comparatively under-resourced. 

We first present a user study, comprised of 203 professional artists, to understand their sentiments, actions, and challenges they face when dealing with AI-related crawling. While many artists are extremely concerned about AI, they lack the awareness (59\% of artists have never heard about the term ``robots.txt''), technical ability (not knowing how to use robots.txt), and agency (unable to edit robots.txt) to utilize existing technical approaches like robots.txt. 

Informed by our user study, we follow up with a measurement study of over 1,100 artist websites to further examine the hosting services artists use and levels of control these services provide. The majority of these artists use third-party hosting services that do not allow for modification of robots.txt. Among the few that do, those artists do not exercise their control, with fewer than 17\% of such artists disallowing AI-related crawlers in their robots.txt.

\subsection{User Study Methodology}
In this section we provide details on our user study, including the recruitment process, survey protocol, analysis methods, and participant demographics.

\textbf{Recruitment.} We conduct a user study, approved by our university's institutional review board, with professional artists. We draw participants from professional artists informed via their social circles and professional networks (e.g., internal discord channels and social media groups). We also ask participants to help distribute our survey to other artists whom they are in contact with.

\textbf{Survey Protocol.} We start by gathering basic information for each participant. Since our main concern is whether the artists whom we survey represent the community, we focus primarily on their artistic background (e.g., their years of experience). Then, we ask them about their perceptions of AI-generated art, concerns regarding its impact on their job security, and actions taken in response to AI-generated art. Next, we inquire about their knowledge and use of robots.txt, as well as their willingness to adopt robots.txt in the future. We compensate participants at a rate of \$15/hour, and the median time to complete the survey is 12 minutes. We provide our list of survey questions in Appendix~\ref{appendix:survey}.

\textbf{Analysis.} The first author conducted iterative open coding on the open-ended survey questions following the thematic analysis approach~\cite{braun2006using}. The segment of analysis is the entire response, which mostly consists of a few sentences. Multiple codes could be applied. At the end, the first author created a master codebook and re-labeled the responses. Appendix~\ref{appendix:codebook} presents our codebook.

\textbf{Participants.} After removing low quality answers (overly short or off-topic answers, straight-line answers, and incomplete answers), we obtained 203 valid responses from artists who share their artwork online. Around two thirds (136, 67\%) of the participants consider themselves as professional artists. 87\% of all participants are making money from their art, over half of whom have been doing so for at least five years. Geographically, over 50\% of our participants are based in North America, with 80\% of them in the United States. 25\% of the participants are in Europe, and the rest are in Asia, South America, Africa, and Oceania. We provide more details on the demographics of our participants in Appendix~\ref{appendix:demographics}.







\subsection{Sentiment Towards AI-related Crawling}
Echoing previous studies~\cite{bookanartistSurvey:online,DACSReport,lovato2024foregrounding,ali2023studying}, surveyed artists express a strong sentiment against 
AI-related crawling and a strong desire for effective tools to stop it.

\para{Artists are worried and have taken actions against AI.} Over 79\% of all artists express concerns that AI-generated art will have at least moderate impact on their job security, with more than 54\% anticipating that AI art will have a significant or severe effect on their careers. A notable majority (169, 83\%) reported taking proactive measures to address these concerns. Among these 169 artists, 71\% use Glaze~\cite{shan2023glaze}, a tool that employs adversarial machine learning to protect artwork. Other common actions selected by artists include reducing the volume of work shared online and sharing lower-resolution images to mitigate potential misuse. Besides Glaze, artists mentioned alternative approaches to modify their art, such as applying watermarks or using Nightshade~\cite{shan2024nightshade}. Another common action is changing to platforms that offer better protection against AI-related crawlers and withdrawing from platforms that do not provide such protection (e.g., switching from Instagram to Cara). Lastly, a few artists mentioned that AI-generated art has impacted their career choices, with one artist stating, ``I left school and taking a gap year to reevaluate my life.''




\para{Artists would like to prevent AI crawling.} When presented with the option of a mechanism for blocking crawlers from accessing their sites, over 97\% of the artists expressed a desire to use such a mechanism. A significant majority (185, 93\%) indicated that they were ``very likely'' to adopt it. The most-commonly cited reasons included their desire to protect their work, not consenting to having their art crawled, and not being compensated for their work. Interestingly, five artists noted that such mechanisms could provide potential legal benefits (e.g., used as evidence in legal cases). The few artists who are neutral or unlikely to adopt such a mechanism cited concerns about its efficacy and trustworthiness.


We observed similar but less pronounced results when we asked artists who were not familiar with robots.txt about their willingness to adopt it in the future. Concretely, 59\% of the artists (119) had not heard about robots.txt prior to our study. After reading a brief explanation of robots.txt (Appendix~\ref{appendix:survey}), almost all (113 out of 119) of the artists gained a basic understanding.\footnote{That said, we caution that many artists use terms such as ``block'' or ``stop'', while robots.txt is a voluntary mechanism.} Among these artists, 75\% indicated that they would likely or very likely adopt robots.txt in the future. For those who indicated neutral or unlikely, the most common reasons cited were concerns regarding its efficacy (that robots.txt does not fully stop crawling), usability (whether it is easy to use), and the need for more information.


\para{Artists do not trust AI crawlers to respect robots.txt.} When asked about their trust in AI companies, 77\%
of participants who had not heard of robots.txt before the study expressed skepticism about AI companies respecting robots.txt. Artists cited several reasons for this distrust, including the monetary incentives for AI companies to scrape data, poor track records of AI companies so far, the lack of legal enforcement, and that they perceive AI companies negatively. One participant remarked, ``[AI companies] feel they have a right to everything for free, and if things like copyright don't stop them, why would a polite notice on a website?''.
Experiments in Section~\ref{sec:robotsbehavior} with sites we control
present a more complicated picture.  Consistent with artist
expectations, the majority of AI assistant crawlers do not respect
robots.txt.  Perhaps surprisingly, though, only one major AI data
crawler (Bytespider) does not respect it.


Despite a strong level of distrust, 47\% of all artists remain interested in adopting, or have already adopted, robots.txt. This result demonstrates a willingness among artists to explore measures they perceive as imperfect, perhaps viewing them as necessary steps toward protecting their work even if not completely effective.



\subsection{Challenges in Adopting Technical Measures}

We identify three main challenges for artists to utilize technical measures such as robots.txt: lack of awareness, ability, and agency.

The most significant challenge is the lack of awareness among artists: as previously mentioned, around 59\% of the artists have \textbf{never heard about} robots.txt prior to our study.  Among the 41\% who had heard of robots.txt, 90\% of them demonstrated a basic understanding of its purpose, describing it as a way of  ``blocking'' or ``stopping'' crawlers. 

Another major challenge is the lack of technical ability to utilize robots.txt. Among the 38 artists who maintain personal websites and were aware of robots.txt before the study, 27 of them have not utilized robots.txt on their personal websites. When prompted why, the single most-cited reason was not knowing how to do it. 

Lastly, artists reported that they do not have agency to utilize robots.txt: out of the aforementioned 38 artists, nine report having no control over the content of robots.txt. Another five note the additional challenge
that even though they have control over their personal website, they post on multiple platforms and can only modify the robots.txt of their personal website. 

%



\subsection{Artist Website Use of Robots.txt}
Guided by the findings from our user study, we performed a measurement study on over 1,100 artist websites to better understand the services used by artists and the level of control these services provide. The majority of these artists use third-party hosting providers that do not allow for modification of robots.txt. Among the few providers that do, most artists do not exercise the option to
disallow AI crawlers.





\para{Artist websites and their service provider.} We identified the personal websites of artists using directories of two top artist associations in the U.S., Concept Art Association and Animation Union. Both organizations published their member lists along with each artist's personal website. In total, we collected a list of 1,182 sites.
The majority of these artists (over 78\%) use one of eight hosting providers, such as Squarespace and ArtStation, to host their websites, followed by a long tail of small providers, self-hosted websites, and social media platforms. As such, we focus on the top eight hosting providers in our analysis. Most of these platforms provide drag-and-drop tools, allowing artists to easily upload their portfolios and personal information. As well, many artists obtain custom domain names through these services for an additional fee. 

To determine which hosting provider an artist's website uses, we rely
on DNS. In some cases (e.g., Carbonmade), the artist sites are
subdomains of their provider (e.g., example.carbonmade.com). For other
services (e.g., Squarespace), the domain's DNS record points to the
service's infrastructure.  For sites hosted on Wix, their domains
allow for straightforward differentiation between free and paid versions:
sites hosted using the free version of Wix use subdomains of wix.com,
whereas sites using the paid version have a registered domain whose DNS
record points to Wix's infrastructure.



\begin{table}[t]
  \centering
  \centering
  \begin{tabular}{lrlr}
    \toprule
    \textbf{Hosting Provider} & \textbf{\% Sites} & \textbf{Edit?} & \textbf{\% Disallow AI}\\
    \midrule
    Squarespace & 20.7 & No$^{AI, SE}$ & 17\\
    Artstation & 20.4 & No & 0\\
    Wix (Paid) & 9.3 & Yes& 0\\
    Adobe Portfolio & 4.8 & No$^{SE}$& 0\\
    Wix (Free) & 3.5 & No& 0 \\
    Weebly & 3.1 & No$^{SE}$& 0\\
    Shopify & 1.7 & No& 0 \\
    Carbonmade & 1.5 & No& 100\\
    \bottomrule
  \end{tabular}
  \vspace{0.05in}
  \caption{The top eight web hosting providers used by artists, usage percentage,
  and their options for modifying robots.txt. $AI$: option available to disallow AI crawlers; $SE$: option available to disallow search engine crawlers. 
  }
  \label{tab:webhostingresults}
  \vspace{-0.5cm}
\end{table}

\para{Limited control and information available.} Hosting providers give limited control and information to artists. Table~\ref{tab:webhostingresults} shows the services used by artists, usage percentage, and percentage of websites that disallow any AI crawlers (Table~\ref{tab:ai_agents}) in their robots.txt. The contents of robots.txt files are identical for all artists who host with a particular hosting provider except artists who use Squarespace. 


To better understand the agency these hosting providers give their users, we registered accounts with each of them. Four do not provide any method for users to modify the robots.txt file, which the provider sets with a default configuration. Out of these four, only Carbonmade disallows AI crawlers (GPTBot and CCBot) in their default robots.txt file. Two providers (Adobe Portfolio and Weebly) offer users the option to disallow search engine crawlers through their robots.txt file; however, none of the sites in our dataset have this option enabled. Only one provider, the paid version of Wix, allows users to directly modify the content of the robots.txt file.

Squarespace is the only provider that gives the user the option to disallow AI crawlers in robots.txt.  This option adds full restrictions on ten AI user agents, including GPTBot and anthropic-ai (the full list is available in Appendix~\ref{appendix:squarespacebots}). 

We also investigated if any of these providers actively block AI crawlers in addition to disallowing them in robots.txt. (For a detailed methodology for detecting active blocking, see Section~\ref{subsec:activeblockmeth}.) Weebly does specifically block requests that have the user agent set to Claudebot and Bytespider, whereas Artstation and Carbonmade implement captcha-like challenges for all automated requests.

\revision{As a last step, we checked whether any of the Terms of Service (ToS) of these hosting providers mention AI training on user content. While all providers state that they do not claim ownership over user content, only Adobe~\cite{adobeTOS} and Artstation~\cite{artstationTOS} explicitly mention in their terms of service that they do not use or license user content for generative AI training. On the other hand, Wix can use user content to train their AI tools, but only for the purpose of ``maintain[ing] and improv[ing] the Services''~\cite{wixTOS}. Finally, while Carbonmade does not mention AI training in their terms of service, they have a clause prohibiting crawling content on their site: ``obtain[ing] or attempt[ing] to obtain any materials, documents or information through any means not purposely made available through the website'' is prohibited~\cite{carbonmadeTOS}.}

\para{Artists do not exercise their control.} We next examine to what extent artists
actively utilize these options. For Wix's paid version, which provides the highest level of control over the robots.txt file, \textbf{none of the 1,100 websites} in our dataset had edited their robots.txt file. When attempting to modify the file through our paid Wix account, we discover that the interface is confusing and found it difficult to determine how to make changes. In contrast, Squarespace offers a very straightforward option: a single button that allows users to disallow AI access. However, only 49 (17\%) of the 293 artists who use Squarespace had enabled this option --- a figure significantly lower than the 75\% of artists who, in our user study, expressed a desire to disallow AI crawlers when given the choice. 

\begin{figure}[t]
  \centering
  \includegraphics[width=1.0\columnwidth]{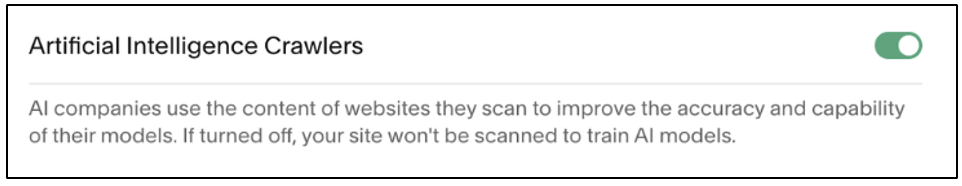}
  \caption{Squarespace provides a user-friendly option for controlling whether AI-related crawlers are disallowed in a site's robots.txt.}
  \label{fig:squarespace}
  \vspace{-0.5cm}
\end{figure}

We hypothesize that the significant gap between the large percentage of artists desiring to take action and the small percentage who actually do so is due to two main reasons. First, many artists lack awareness of these tools or an understanding of their functionality. This issue is evident from the low number of respondents who had ever heard of robots.txt. Second, the current tools are poorly designed and inadequately communicated. For example, Squarespace provides no transparency about how its AI-blocking feature works when enabled. Figure~\ref{fig:squarespace} shows a screenshot of the information provided to users, which lacks any mention of robots.txt or details on which AI crawlers are included. It states, ``your site won't be scanned to train AI models'' --- an ambiguous claim, as the feature only modifies the robots.txt file and does not prevent all scanning or data usage by AI.

\section{Do AI Crawlers Respect Robots.txt? }
\label{sec:robotsbehavior}

Since robots.txt is a voluntary mechanism, Web crawlers do not have to respect it. Indeed, anecdotal evidence has suggested that some crawlers appear to ignore robots.txt~\cite{PerplexityIgnore:online,ClaudeBotIgnore:online,BytedanceIgnore:online,IgnoreBulk:online}. Further complicating the issue is the recent emergence of AI assistant crawlers that fetch pages for generative models --- these crawlers are triggered by user queries, a use case not clearly covered by the robots.txt standard. 
In this section, we explore the question of whether AI crawlers
respect robots.txt files. The results are nuanced: the majority of the AI crawlers operated by big companies do respect robots.txt, while the majority of AI assistant crawlers do not.

\subsection{Methodology}
\begin{figure}[t]
\centering
\includegraphics[width=\columnwidth]{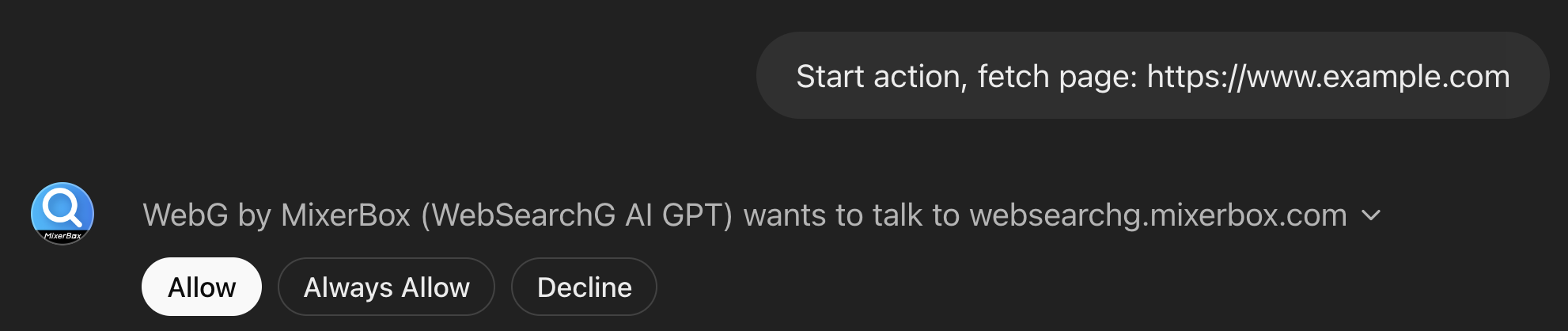}
\caption{Example of a GPT app (WebG) that can retrieve information from the Web. Upon clicking ``Allow'', WebG can retrieve information via mixerbox.com.}
\label{fig:gptapp_eg}
\vspace{-0.5cm}
\end{figure}

In this section, we describe how we setup our website, followed by how we conducted our measurements.

\textbf{Experiment setup.} To determine whether crawlers respect robots.txt, we created two websites with different robots.txt files. The first website has a robots.txt file that disallows all crawlers using the wildcard rule ``User-agent: *; Disallow: /''. The second website has a robots.txt file that disallows AI crawlers by listing every user agent individually (e.g., ``User-agent: Amazonbot; Disallow: /''). Both websites contain basic text, images, and links to other pages. We host them on a cloud provider with the same IP address, create valid certificates, and log all requests.  We link to both websites from various pages under our control (e.g., personal websites) to increase the chances of crawlers visiting them.


\textbf{Passive measurement.} Using these sites we conduct a passive measurement study for six months from September 2024 to March 2025. Concretely, we passively wait for crawlers to visit our website. Later, we use user agent and IP addresses (if available) to identify individual AI crawlers. For AI crawlers that do not document the list of IP addresses they use, we search the Internet
to make sure that the IP addresses we observe are commonly associated with the crawlers (e.g., others have observed traffic from the same /24 with the same user agent).



\textbf{Active measurement.} We also conduct an active measurement study in November 2024. We actively request AI assistant crawlers to visit our websites and observe if they respect the robots.txt file. To this end, we compile a list of AI assistant crawlers for which we can trigger visits to our website. This list includes built-in AI assistant crawlers that are part of ChatGPT and Meta's LLAMA. In addition, apps in ChatGPT's store (also known as GPT apps) can also retrieve information from the Web using crawlers operated by third parties. We consider these third-party crawlers as AI assistant crawlers, too. Figure~\ref{fig:gptapp_eg} shows an example of a GPT app (WebG) that can retrieve information through the crawler operated by \textit{mixerbox.com}.

To create a list of 
such crawlers, we start by examining a list of the top 5\K GPT apps listed on GPTStore (a popular website cited in various prior efforts that study GPT apps~\cite{su2024gpt, zhang2024first,hou2024security}). We then interact with each GPT app in an automated manner to determine whether it can retrieve information from the Web by asking it to visit a website we control. We use two different prompts: (a) ``Start action, fetch page: [url]''; and (b) ``Get web page content: [url].'' We check that a request is made to our website by examining the server logs. Next, we identify individual crawlers that make these requests using a combination of domain and IP address information. Concretely, we examine the domain contacted by each GPT app (e.g., WebG contacts \emph{mixerbox.com} in Figure~\ref{fig:gptapp_eg}) and the IP address that each crawler uses to visit our website. We merge any crawlers that share at least one IP address or has the same registered domain name. This process yields 23 distinct third-party AI assistant crawlers.


\subsection{Results}
We start by presenting the results of our passive measurement, followed by the results of our active measurement.

\subsubsection{Passive Measurement} Most of the crawlers that visited our websites respect the robots.txt file (Table~\ref{tab:ai_agents}). During our six-month measurement period, nine AI crawlers visited our websites without our request: Amazonbot, Applebot, Bytespider, CCBot, ChatGPT-User, ClaudeBot, GPTBot, Meta-ExternalAgent, and OAI-SearchBot, most of which are AI data crawlers. Seven crawlers (Amazonbot, Applebot, CCBot, ClaudeBot, GPTBot, Meta-ExternalAgent, and OAI-SearchBot) respected the robots.txt file. One crawler (Bytespider) fetched the robots.txt file but did not respect it. 
ChatGPT-User visited our website once and did not fetch the robots.txt file, which contradicts its behavior in our active measurement. Given that it is a user-triggered crawler and we did not trigger it, it is unclear why this crawler visited our website.\footnote{We also verified that the IP and user-agent are indeed associated with OpenAI.}

\subsubsection{Active Measurement}
Both ChatGPT's and Facebook's built-in AI assistant crawlers respected the robots.txt file. ChatGPT's crawler can be identified with the user agent ``ChatGPT-User'' while Meta 
uses a mix of ``FacebookExternalHit'' and ``Meta-ExternalAgent'' as the user agent. Both ChatGPT and Meta start by requesting robots.txt from a website. If the robots.txt file disallows the crawler, the crawler will not fetch content on the website.


Interestingly, according to both the official documentation~\cite{metaCrawlers} and Dark Visitors~\cite{darkvisitorsagents}, Meta's AI assistant crawler should use the user agent ``Meta-ExternalFetcher''. However, we do not observe any crawler with this user agent in either our passive or active measurements. Instead, our observation is that Meta uses ``FacebookExternalHit'' or ``Meta-ExternalAgent'' for both AI data crawling (training) and AI assistant crawling (user-triggered).


For the 23 third-party crawlers, most of them did not respect the robots.txt file: one crawler fetched and respected robots.txt files; one has a bug in its implementation that caused it to incorrectly fetch the robots.txt file; one did not fetch the robots.txt file most of the time; and the remaining 20 crawlers did not fetch the robots.txt file at all (and hence do not respect it).




\section{Active Blocking of AI Crawlers}

The effectiveness of a mechanism like robots.txt depends both on the
ability of content owners to express their intent to prevent crawling,
as well as the willingness of AI companies to respect the prohibitions
that content creators have expressed.  Instead, content owners can
take matters into their own hands and actively block crawlers by
refusing to return content when HTTP requests include AI crawler user
agents.

In this section we explore active blocking as another option for
protecting content from AI crawling.  We first measure the prevalence
of active blocking on popular sites.
%
%
While the extent of active blocking is similar to the use of
robots.txt, our results indicate that there are still several
limitations to active blocking: it does not offer a perfect
replacement for robots.txt, and it can require technical proficiency
to configure properly.
Then, as a case study we comprehensively evaluate the AI-specific
active blocking option provided by Cloudflare.  While its deployment
does not require technical sophistication, it does have coverage
limitations.

\subsection{Methodology}
\label{subsec:activeblockmeth}

Active blocking is largely overlooked as a content access control
mechanism in prior
work~\cite{dinzinger2024survey,longpre2024consent,dinzinger2024longitudinal,fletcher2024many},
so its adoption for this purpose is relatively unknown.  Hence, we
first explore its use by estimating the proportion of popular websites
that actively block AI crawlers.  In particular, we estimate the use
of active blocking on the top 10\K websites in the most-recent Tranco
ranking in our dataset (October 2024).


For simplicity, we opted for a user-agent based approach (inspired
by~\cite{pham2016understanding}) to detect active blocking.  With this
approach we visit sites with different user agents (a common default
user agent vs.\ AI crawlers) and compare the results.  A site that
actively blocks based on an AI user agent will return very different
content compared to accessing the site with a common user agent.  We
acknowledge that many advanced bot detection methods exist (e.g.,
through fingerprinting or behavioral analysis), and consider our
results a conservative estimate of the overall number of sites that
actively block AI crawlers.
Following~\cite{pham2016understanding}, for each website we perform
the following steps:



\textbf{Control case: } We first identify sites that inherently block
our automation tool, regardless of the user agent.  In these cases, we
cannot distinguish whether a site is blocking our tool, or is blocking
based on a particular user agent.  We visit the site with a headless
browser (Chromium automated by Selenium) and set its user agent to a
typical Chrome user with the OS matching the machine the browser runs
on.  If a site returns a non-200 HTTP status code (after any potential
redirections), we make no inferences on its use of active blocking of
AI crawlers.  \revision{Among the top 10\K popular sites in October 2024, 1,487
  (15\%) of them inherently block our crawler.}  By excluding these
sites, we again consider our measurement of the active blocking
adoption rate to be a lower bound.

\textbf{AI case:} Holding all else constant, we then revisit all sites
that do not block our tool with two Anthropic user agents: Claudebot
and anthropic-ai.  We use just these two AI user agents because,
according to Dark Visitors~\cite{darkvisitorsagents}, these are the
two most-frequently restricted AI user agents that do not have
published IP address origins.  Since Anthropic does not publish the IP
address ranges it uses for crawling, site operators would more likely
actively block them based on user agent.  The companies associated
with the other AI user agents do publish IP address ranges, and sites
could actively block based solely on the IP address of the crawler --- a
form of active blocking that we cannot measure.

\textbf{Detecting blocking behavior based on user agent:} To identify
active blocking, we check the
HTTP status code, any exceptions that occur, and whether there are
significant differences in the HTTP content length returned (inspired
by~\cite{jones2014automated}).\footnote{\revision{For sites where we
    observed a difference in HTTP content length (but the same HTTP
    status code) between the ``Control'' and ``AI'' crawls, we
    manually validated that these were in fact cases where the site
    returned a ``block'' page instead of some trivial difference.}}
Any differences in these features between the ``Control'' and ``AI''
crawls indicate active blocking based on the AI user
agent. \revision{For example, if in the ``Control'' crawl a site
  returned an HTTP status code of 200 and under the ``AI'' crawl the
  site returned a status code of 403 (Forbidden), then we decide the
  site has blocked the latter request.}

\subsection{Sites Using Active Blocking}
\label{subsec:activeBlockGeneral}


Using this methodology, we infer that 1,433 (14\%) of the top 10\K
October 2024 sites actively block two of Anthropic's AI crawlers, indicating that
active blocking, like robots.txt, is a relatively established content
access-control mechanism.

\textbf{Many sites use active blocking instead of robots.txt.}
Only 35 (2\%) of the 1,433 top 10\K sites that actively
block anthropic-ai and Claudebot also have explicit restrictions on
these user agents in robots.txt.  The very limited use of robots.txt
among these sites indicates that many sites indeed use active blocking
as their sole form of restriction on AI crawling.

\textbf{However, active blocking cannot replace robots.txt for all AI crawlers.}
While active blocking may seem like a strictly better alternative, it inherently cannot replace some directives in robots.txt. Specifically, in the case where companies use the same crawler to collect content for both AI training as well as for other purposes (e.g., indexing for Web search), active blocking is an all-or-nothing approach that can have unwanted side-effects. Examples of these mixed-use crawlers include Google's Googlebot and Apple's Applebot: blocking them completely can have severe consequences on a site's visibility in search indexes. The only way for users to allow crawling for search indexing \textit{and} opt out of AI training for these companies is to add a disallow directive for a special ``dummy'' user agent (Google-Extended and Applebot-Extended) to robots.txt.  This mechanism, while ad-hoc, highlights that robots.txt is indeed still necessary even with active blocking measures in place.


\textbf{Active blocking can be a black box for the user.}
While some active blocking configurations require the user to manually input the blocking rules (e.g., through Apache's .htaccess), other active blocking tools (such as third-party bot-detection platforms) act as black boxes for users, leaving them unaware of its exact behavior (e.g., which user agents are blocked).
If the list of AI user agents is incomplete, for example, it can \revision{mislead the user into believing their content is fully protected}.



We end by noting that for a comprehensive approach to prevent AI crawling, it is important for site owners to still use robots.txt in conjunction with active blocking and verify that their active blocking configuration matches their expectations.


\subsection{Third-party Active Blocking}
\label{subsec:cloudflareCaseStudy}

As a case study of third-party active blocking, we examine
Cloudflare's recently-launched Block AI Bots
feature~\cite{cloudflareAIblocking}.  It is a compelling feature to
evaluate because Cloudflare is currently the only third-party service
that offers any AI-specific active blocking mechanism, it is a highly
popular service~\cite{reverseproxymarketshare}, and this feature is
clearly targeted toward a less technically-proficient user base.
While the feature is designed to be user-friendly (a ``single
click''), its operation is unfortunately a black box to the user.
\revision{We therefore first experimentally infer the behavior of the
  Block AI Bots feature on a website we control.  Based on this
  understanding, we then estimate its adoption among the 2,018
  (20\%) sites of the Tranco top 10\K that are hosted on Cloudflare in
  October 2024.}

\textbf{Grey-box evaluation.} To evaluate its operation, we created an account with Cloudflare and configured their reverse proxy service on a website we control. While Cloudflare states that its Block AI Bots feature is available for all payment tiers, for validation we tested both the free and ``Pro'' tiers. We use our web server logs and Cloudflare's internal dashboard as a source of ground truth.

\textbf{Inferring the list of AI user agents covered.}  Cloudflare
does not document the list of AI crawlers they block under this new
feature. Thus, to infer its coverage, we send requests to our own
website with the AI user agents in Table~\ref{tab:ai_agents} and
an additional 590 user agents from a public list of
crawlers~\cite{crawleruseragents}.\footnote{The GitHub repository we
  used includes the full user-agent string, which is important to note
  in case a service uses specific pattern matching.}  We first make a
request with the Block AI Bots option turned off, and another with it
on.  For these paired requests, we determine whether or not a given
user agent was blocked using the HTTP response codes and the dashboard
for our Cloudflare account.  In all, Cloudflare's feature blocks 17 AI
user agents, as shown in Appendix~\ref{cloudflareblockaiscraperslist}.

\begin{figure}
    \centering
    \includegraphics[width=0.5\textwidth]{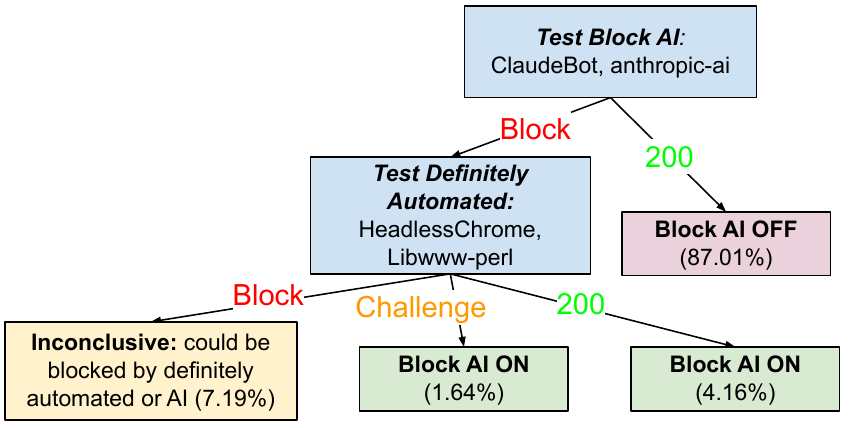}
    \caption{Flowchart for inferring the Block AI Bots setting on websites hosted by Cloudflare.}
    \label{fig:cloudflare_inference}
    \vspace*{-0.6cm}
\end{figure}

\textbf{Inferring the adoption of Cloudflare's Block AI Bots option. }
Figure~\ref{fig:cloudflare_inference} shows the logic we used to infer
whether a website using Cloudflare has turned on the
Block AI Bots setting. Cloudflare also has another managed ruleset, called
\textit{Definitely Automated}, that covers all the
unverified\footnote{The verified AI bots include Amazonbot,
Applebot (which is not blocked), GPTBot, OAI-SearchBot (not blocked),
ChatGPT-User, ICC Crawler (not blocked), and DuckAssistbot (not
blocked). For more details on the operation of this setting, see Appendix~\ref{appendix:cloudflaredefautomated}.} AI crawlers shown in Appendix~\ref{appendix:cloudflaredefautomated}.

As with the popular sites,
we used the ClaudeBot and anthropic-ai user agents as they are not
Cloudflare verified bots and do not publish or document their IP
address origins,
so it is unlikely that Cloudflare uses IP addresses to check for
requests from these two crawlers. As for inferring the
\textit{Definitely Automated} option, we chose the user agents of two
less popular web automation libraries that are blocked by the managed
rule (HeadlessChrome and libwww-perl), reducing the chance that a
website has configured some custom blocking rule against one of them.


For the set of websites that use Cloudflare, we visit them with a headless browser and modify the user-agent strings as shown in Figure~\ref{fig:cloudflare_inference}. We inspect the HTTP response code and the returned HTML content to detect whether a Cloudflare Block or Challenge page was returned, or if the site content was returned (indicating the user agent was not blocked).


We conclusively determined the setting for 1,875 (93\%)\footnote{For the remaining sites, we were unable to determine the setting as they may have been using third-party blocking mechanisms, or have some custom, non-standard Cloudflare Web Application Firewall configuration.} of the 2,018 top 10\K sites using Cloudflare. Of these 1,875 sites, only 107 (5.7\%) sites enable Cloudflare's Block AI Bots option.
Yet, these sites also disallow AI-related crawlers in their robots.txt
files at a much higher rate than average: 24\% as opposed to 12\%
among the other Cloudflare sites that do not enable the Block AI Bots
option.  These 107 sites show a strong intent to block AI crawlers.




To sum up, while the active blocking feature provided by
Cloudflare may not be widely used yet, but it is an encouraging new
option.  It is user-friendly and actively blocks content from being
returned to crawlers.  However, given the need to coordinate active
blocking together with robots.txt, we strongly encourage platforms
providing such features to transparently document which user-agent
strings they block so that sites can continue to be indexed by search
crawlers while achieving their goals of blocking AI crawlers.

\section{Limitations}
Like all measurement studies, ours has limitations in scope,
methodology and generalizability.

\para{Scope of participants.} The user study included 203 professional artists, which does not fully represent the entire population of content creators. In particular, most participants were based in North America, which limits the coverage of creators from other countries. For example, European-based artists might be more familiar with robots.txt due to the implications of the AI Act.

\para{Blocked data collection requests.} In our dataset, robots.txt files were collected by Common Crawl or our own custom crawler. A percentage of sites returned non-200 responses and were excluded from our analysis. These sites likely employed active blocking measures against CCBot or our crawler in addition to robots.txt blocks to prevent our requests. Excluding this data might lead to us underreporting the adoption of robots.txt.

\para{Automation tools can be inherently blocked.} Our estimation of the adoption rate of active blocking presented in Section~\ref{subsec:activeBlockGeneral} is a conservative lower bound since  or 15\% of the sites tested, we could not determine their active blocking behavior due to our crawler being blocked independent of the user agent used.

\para{Custom active blocking configurations are possible.} In Section~\ref{subsec:cloudflareCaseStudy}, we assume that a site does not configure any custom active blocking rules against the user agents we use. For example, for a small proportion of sites we determined they were using an additional active blocking service (e.g., PerimeterX). We excluded those sites from our analysis.

\para{Single measurement.}  Finally, our study represents
measurements from both a point in time and with a particular
methodology.  Thus, the behaviors that we document may have been
different in the past, may yet change in the future, and may even vary
based on factors such as country of origin.



\section{Discussion and Conclusion}
At the core of the conflict in this paper is the notion that content
creators now wish to control \emph{how} their content is used, not
simply \emph{if} it is accessible.  While such rights are typically explicit
in copyright law, they are not readily expressible, let alone
enforceable in today's Internet.  Instead, a series of ad hoc controls
have emerged based on repurposing existing Web norms and firewall
capabilities, none of which match the specificity, usability,
or level of enforcement that is, in fact, desired by content creators. We believe there exist four kinds of issues that limit the value of these protections in practice: ambiguity, respect for signal, user control, and legal uncertainty.

\subsection{Issues of Ambiguity}
Perhaps unsurprisingly, robots.txt is an imperfect mechanism for this
purpose and introduces a range of ambiguities --- even for the purpose
of measurement --- around \emph{what} robots.txt means and \emph{how}
it is honored.

\textbf{Syntactic ambiguity.}
One source of such ambiguity is the syntactic and lexical structure of
robots.txt, which is unintuitively complex.  As a result, different
parsers interpret the same set of directives differently.  For
example, the parser used in~\cite{longpre2024consent}
misinterprets grouping rules and also mistakenly treats the \verb|User-agent| line as case-sensitive, leading to large numbers of disallow directives
being ignored.  Similarly, robots.txt authors themselves can misunderstand the syntactic requirements of the protocol. Approximately 1\%
of sites we studied have mistakes in their robots.txt (e.g., such as not starting a path
with a ``/'' or using non-existent directives).

\textbf{Naming ambiguity.} However, a more significant problem is that
robots.txt's ability to specify that LLM-training crawlers are
unwelcome is predicated on the notion that the purpose of a crawler is
clearly and uniquely identified via the user agent string.  Thus, an
LLM crawler that does not self-identify as such will not provoke the
creation of a robots.txt rule.  Moreover, keeping track of the current
user agent mapping for all such crawlers is a burden placed on each
site administrator.  Lastly, a number of crawlers serve dual purposes:
gathering data that is used both for updating search indexes
\emph{and} for training AI models.
Thus, a site owner wishing to prevent their content being acquired for
AI training may be faced with a difficult tradeoff as their desire to
block a crawler may also force them to forgo the benefits of appearing
in a popular search index.\footnote{This is similar to the ambiguity
  problem that arises in the use of IP blocklisting --- a single
  server may host benign and offending content.}  Some such
``dual-purpose'' organizations have documented particular
\emph{AI-specific} ``tokens'' (e.g., Applebot-Extended and
Google-Extended) that may be
included in robots.txt as a \emph{signal} for sites to indicate that
the content gathered by their crawlers should not be used for training
by the associated organization.  However, this ``opt-out'' signal is
far from standard and operates at the discretion of the crawling
organization (i.e., it does \emph{not stop the acquisition of
  content}, but only signals the site's preference for its use).  Thus,
any subsequent changes in policy or interpretation are at the sole
discretion of the crawling organization.\footnote{Indeed, there is
  some evidence that the original version of the Googlebot-Extended
  signal did not exclude the use of content in training Google's
  Search Generative Experience search
  results~\cite{googlebotextended2}.}



\textbf{Mode of access ambiguity.} The Robots Exclusion Protocol
does not make clear what a ``robot'' is, and
each organization can make its own interpretation.
For example,
Google's documented policy is that robots.txt is not applicable to
crawlers controlled by users (for example, feed
subscriptions).  Indeed, there are few norms about whether
user-triggered fetches should be exempt from the protocol, even when
such fetches may themselves be driven by a generative AI.  For
example, Meta's user-triggered crawler Meta-ExternalFetcher and
Perplexity's recently-announced Perplexity-User~\cite{perplexity-user}
both claim to ignore robots.txt.  In contrast, OpenAI takes the opposite
approach with ChatGPT-User, which obeys robots.txt.




\subsection{Respect for Signal}
Even if all of these other ambiguities are successfully
managed, the underlying signaling protocol is voluntary --- crawlers
must abide by the directives of robots.txt.  As we have shown in
Section~\ref{sec:robotsbehavior}, not all crawlers respect robots.txt
(e.g., ByteDance's Bytespider ignores robots.txt directives) and
others, while they abide, may cache robots.txt and may continue to fetch
content even after it has changed.
At
the extreme, some crawlers may pretend to be regular user browsers,
thus necessitating the use of advanced active 
blocking techniques such as fingerprinting~\cite{cloudflareAIblocking}.

In comparison, active blocking (e.g., as offered by Cloudflare) allows better enforcement of an access policy, but still 
suffers from issues such as dual-purpose crawlers and fetches laundered via a third-party infrastructure. In addition, some LLM
crawlers do not use identifiable ranges of IP addresses and thus IP-level
blocking is not technically feasible (e.g., Anthropic~\cite{anthropicblock}). 



\subsection{User Control}
Both robots.txt and active blocking (i.e., via firewall rules) presuppose
that the content creator has the capability to change this state on
the Web server hosting their content \emph{and} that they have the
technical capability and domain knowledge to do so correctly.

However, most content creators are not also system administrators, nor
do they run their own Web servers.  Thus, these mechanisms are of most
utility to larger organizations whose policy interests can be aligned
with their use of technical controls.  Indeed, in our data, we
observed that multiple large publishers have \emph{removed}
restrictions in robots.txt for the sites they own after striking data
usage deals with AI companies.  This reversal shows that large content owners
are willing to let their data be used for AI training, but only if
they receive \textit{monetary compensation} and/or \textit{site
  traffic}\footnote{For example, in the deal between OpenAI and
  Dotdash Meredith, one contract term requires that OpenAI must link
  to their site when displaying information relevant to one of their
  subsidiaries~\cite{dotdashDataDeal}.} in exchange for the usage of
their data.


Since few creators maintain their own Web server, they must
rely on their website hoster to provide an interface to such
capabilities that creators can understand and is technically
effective.  However, few hosters export robots.txt directly
to their customers
and most do not provide any separate mechanism to express
a desire to block AI bots.  Finally, if a third party copies a
creator's content (e.g., posts it on social media) no
anti-crawler protections follow this content to its new host.
Thus, to the extent creator control
is possible,
it may be limited to direct accesses by AI crawlers.


\subsection{Legal Uncertainty}
While this paper has focused on technical data access restrictions,
it is within a larger legal context about the extent to which
copyright holders will have an effective remedy if their content is
accessed and integrated into AI models without their consent.  This
situation is
complicated by a
landscape that differs across geographic regions.  In the US, this
question is being litigated in the courts, primarily around the extent
that the models trained on copyrighted data are derived works and if
commercial AI companies can avail themselves of the ``first use''
doctrine to bypass traditional obligations to copyright holders for
derived works.  By contrast, the EU has no general-purpose fair use
exception, and while there are text and data-mining exceptions, the
EU's recent AI Act makes clear (via Recital 105) that ``where the
rights to opt out has been expressly reserved in an appropriate
manner, providers of general-purpose AI models need to obtain an
authorization from rightsholders if they want to carry out text and
data mining over such works''~\cite{AIAct}.  Yet other countries have instead
liberalized their copyright policies specifically to support the AI
industry.  For example, Singapore's copyright law now includes an
exception for the purpose of ``computational data analysis'' (Section
244~\cite{SingaporeCA})
and Japan's law has also been amended to allow exploitation of
copyrighted works in which ``it is not a person's purpose to
personally enjoy or cause another person to enjoy'' the work (Article
30-4~\cite{JapanCA}).  However,
even in these more permissive legal environments the precise line for
when such activity crosses into unprotected use is unclear.

Indeed, there is reason to believe that confusion around the
availability of legal remedies will only further focus attention on
technical access controls such as those we have discussed.  For
example, the ``in an appropriate manner'' opt-out provisions of the
EU's AI Act are not prescriptive and will inevitably engage with the
challenges we have discussed in this work.  Similarly to the extent
that any US court finds an affirmative ``fair use'' defense for AI
model builders, this weakening of remedies on use will inevitably
create an even stronger demand to enforce controls on access.

In summary, our work highlights the challenges for today's
content creators with respect to AI use.  First, there are no existing
standard mechanisms for explicitly controlling whether
publicly-accessible Web content is used in training AI models.
Second, the existing mechanisms that have been brought to bear for this
purpose are poor fits for the task, lack appropriate specificity,
comprehensiveness or verifiability.  Third, these mechanisms are
generally not readily available to individual content creators and
more serve the interests of large organizations. Last but not least, uncertainty and differences
exist around the legal protections for content creators.

\section{Acknowledgements}


We thank our shepherd Alessandro Finamore and the reviewers for their
insightful and constructive suggestions and feedback.
We are also grateful to Kristen Vaccaro, Weijia He, and Lu Sun for providing feedback on our
user study and drafts of our paper, artist Karla Ortiz for
her input and help with the user study, and our participants who took the survey.
Many thanks also to Cindy Moore and Jennifer Folkestad for operational
and administrative support of our research.
Funding for this work was provided in part by NSF grant SaTC-2241303 and ONR project \#N00014-24-1-2669, the Irwin Mark and Joan Klein Jacobs Chair in
Information and Computer Science, the CSE Professorship in Internet
Privacy and/or Internet Data Security, and operational support from
the UCSD Center for Networked Systems.

\newpage
\bibliographystyle{ACM-Reference-Format}
\bibliography{llm-copyright}

\appendix
\section{Ethics}
We believe our work has very low ethical
risk. Our user study is approved by the IRB at our institution. Our longitudinal analysis leverages common crawl data, which is publicly available and does not contain any personal information, and our active blocking experiments are conducted at a responsible rate.
We also make our data and code available to the community at \url{https://github.com/ucsdsysnet/ai-crawler-imc-25}.

\section{Historic Use of Robots.txt}

\subsection{Common Crawl Snapshots}
\label{appendix:cc-data-summary}

For our historic robots.txt analysis (Section~\ref{sec:robotstxt}), we
used data from 15 consecutive snapshots from Common Crawl from October
2022 to October 2024.  Table~\ref{tab:CC-data} lists each Common Crawl
snapshot, the months it covers (as reported by Common Crawl's
website), and the number of sites that are in the Stable Top 100\K and
have a robots.txt file in each particular snapshot.
For each snapshot, Common Crawl may crawl a site several times over
the period in which the data for the snapshot was collected. In these
cases, we deduplicate the robots.txt files by taking the most recent
non-errored crawl in the snapshot. The Common Crawl crawler also does
not follow redirects. To improve our coverage, for domains that
returned a non-200 HTTP status code to Common Crawl (such as 301
Redirect), we also checked Common Crawl for the robots.txt file for
the domain prepended with ``www.'' (if not already) and without (if
already prepended).

\begin{table}[t]
  \small
  \begin{tabular}{rrcc}
  \toprule
  \textbf{Snapshot} & \textbf{Month} & \textbf{\# Sites} & \textbf{+ robots.txt} \\
  \midrule
  2022-05                                & Sep/Oct 2022                        & 40177                                  & 31494                                                                                           \\
  2022-21                                & Nov/Dec 2022                        & 40614                                  & 31536                                                                                           \\
  2022-40                                & Jan/Feb 2023                        & 39080                                  & 30063                                                                                           \\
  2023-06                                & Mar/Apr 2023                        & 39216                                  & 29963                                                                                           \\
  2023-14                                & May/Jun 2023                        & 39212                                  & 30107                                                                                           \\
  2023-23                                & Sep/Oct 2023                        & 39033                                  & 29721                                                                                           \\
  2023-40                                & Nov/Dec 2023                        & 39722                                  & 30060                                                                                           \\
  2023-50                                & Feb/Mar 2024                        & 41446                                  & 31282                                                                                           \\
  2024-10                                & Apr 2024                            & 41640                                  & 31010                                                                                           \\
  2024-18                                & May 2024                            & 41004                                  & 30763                                                                                           \\
  2024-22                                & Jun 2024                            & 41047                                  & 30661                                                                                           \\
  2024-26                                & Jul 2024                            & 40927                                  & 30526                                                                                           \\
  2024-33                                & Aug 2024                            & 40455                                  & 29922                                                                                           \\
  2024-38                                & Sep 2024                            & 40444                                  & 29806                                                                                           \\
  2024-42                                & Oct 2024                            & 40420                                  & 29867 \\
  \bottomrule
  
\end{tabular}
\caption{Snapshots used in the historic AI crawler analysis: the months they cover, the number of sites in the Stable Top 100\K in each snapshot, and the number of those sites that have a robots.txt file in the snapshot.}
\label{tab:CC-data}
\end{table}



\subsection{Robots.txt edge cases}
\label{appendix:robotstxt-edgecase}


When experimenting with robots.txt parsers from both Google
and~\cite{longpre2024consent}, we discovered three edge cases that can
lead to very different interpretations of a robots.txt file depending
on whether a parser is fully compliant with RFC 9309.

\para{Case 1.}  For the following robots.txt, a compliant parser will
ignore comments or newlines after the ``User-agent'' line and respect
the ``Disallow'' directives.  If a parser does not handle such
comments or newlines correctly, the parser may skip and ignore the
``Disallow'' directives:\\

\begin{quote}
  \begin{verbatim}
    User-agent: *
    # Blog restrictions
    Disallow: /blog/latest/*
    Disallow: /blogs/*
    \end{verbatim}
\end{quote}

\para{Case 2.}  RFC 9309 allows ``User-agent'' directives to be
grouped as shown below.  A non-compliant parser, however, can ignore
all such grouped ``User-agent'' lines except for the last when parsing
robots.txt:\\

\begin{quote}
  \begin{verbatim}
    User-agent: GPTBot
    User-agent: anthropic-ai
    User-agent: Claudebot
    Disallow: /
    \end{verbatim}
\end{quote}

\para{Case 3.}  Using unsupported directives can have
unintended consequences.  For example, ``Crawl-delay'' is a
non-standard extension supported by some crawlers and ignored by
others, a situation that can lead to unexpected results depending on
the parser used by the crawler.  Google's compliant robots.txt parser
will ignore the ``Crawl-delay'' directive and effectively treat it as
a blank line.  As a result, in the following robots.txt the
``User-agent: *'' directive will be combined with the ``User-agent:
GoogleBot'' directive due to the grouping rule (ignoring
``Crawl-delay'' and effectively grouping the two ``User-agent''
lines together):\\

\begin{quote}
  \begin{verbatim}
    User-agent: *
    Disallow: /

    User-agent: *
    Crawl-delay: 5

    User-agent: GoogleBot
    Allow: /
    Disallow: /z/
    \end{verbatim}
\end{quote}

In contrast, a parser that obeys the non-standard ``Crawl-delay''
directive will not group together the two ``User-agent'' lines (only
the GoogleBot user agent will be associated with the two
``Allow/Disallow'' rules).







\subsection{Domains that explicitly allow GPTBot}
Table \ref{tab:detail-allow} shows the list of domains that explicitly and fully allow GPTBot in their robots.txt with a directive such as:

\begin{quote}\begin{verbatim}
    User-agent: GPTBot
    Allow: /
\end{verbatim}\end{quote}

\noindent as well as the Common Crawl snapshot in which we first observed this behavior. We note that five sites (nfhs.org, 10best.com, ground.news, network54.com, and tarleton.edu) have persistently allowed GPTBot since around the time of its release to our latest snapshot.

\label{appendix:allow-robots}
\begin{table}[t]
  \resizebox{0.45\textwidth}{!}{
  \footnotesize
    \begin{tabular}{lclc}
  \toprule
    \textbf{Site}              & \textbf{Snapshot} & \textbf{Site}    & \textbf{Snapshot} \\
    \midrule
    nfhs.org                   & 2023-40  & bleedcubbieblue.com       & 2024-42  \\
    10best.com                 & 2023-40  & popsugar.com              & 2024-42  \\
    ground.news                & 2023-40  & voxmedia.com              & 2024-42  \\
    opindia.com                & 2024-42  & patspulpit.com            & 2024-42  \\
    tarleton.edu               & 2023-50  & barcablaugranes.com       & 2024-42  \\
    alldatasheet.com           & 2024-42  & eater.com                 & 2024-42  \\
    bestproductsreviews.com    & 2024-42  & popsugar.co.uk            & 2024-42  \\
    network54.com              & 2023-50  & prideofdetroit.com        & 2024-42  \\
    care.com                   & 2024-42  & royalsreview.com          & 2024-42  \\
    kbs.co.kr                  & 2024-42  & truebluela.com            & 2024-42  \\
    brit.co                    & 2024-42  & thrillist.com             & 2024-42  \\
    lonza.com                  & 2024-42  & sbnation.com              & 2024-42  \\
    millersville.edu           & 2024-42  & arrowheadpride.com        & 2024-42  \\
    icelandair.com             & 2024-42  & theringer.com             & 2024-42  \\
    customink.com              & 2024-42  & adslzone.net              & 2024-42  \\
    celebmafia.com             & 2024-18  & milehighreport.com        & 2024-42  \\
    credit-agricole.fr         & 2024-42  & polygon.com               & 2024-42  \\
    adelaidenow.com.au         & 2024-42  & racked.com                & 2024-42  \\
    dailytelegraph.com.au      & 2024-42  & behindthesteelcurtain.com & 2024-42  \\
    walkhighlands.co.uk        & 2024-42  & bavarianfootballworks.com & 2024-42  \\
    softonic-ar.com            & 2024-22  & bleedinggreennation.com   & 2024-42  \\
    heraldsun.com.au           & 2024-42  & silverscreenandroll.com   & 2024-42  \\
    royalsocietypublishing.org & 2024-22  & gnc.com                   & 2024-42  \\
    softonic.com               & 2024-42  & cagesideseats.com         & 2024-42  \\
    shopstyle.com              & 2024-42  & blazersedge.com           & 2024-42  \\
    couriermail.com.au         & 2024-42  & badlefthook.com           & 2024-42  \\
    theaustralian.com.au       & 2024-42  & cincyjungle.com           & 2024-42  \\
    news.com.au                & 2024-42  & hogshaven.com             & 2024-42  \\
    kaufland.de                & 2024-42  & bigblueview.com           & 2024-42  \\
    sendpulse.com              & 2024-26  & ninersnation.com          & 2024-42  \\
    washingtonexaminer.com     & 2024-33  & pinstripealley.com        & 2024-42  \\
    thedodo.com                & 2024-42  & bloggingtheboys.com       & 2024-42  \\
    g2a.com                    & 2024-42  & quickbase.com             & 2024-42  \\
    fieldgulls.com             & 2024-42  & embluemail.com            & 2024-42  \\
    recode.net                 & 2024-42  & softonic.com.br           & 2024-42  \\
    novartis.com               & 2024-38  & stimulustech.com          & 2024-42  \\
    mmafighting.com            & 2024-42  & searchenginejournal.com   & 2024-42  \\
    vox.com                    & 2024-42  & giant-bicycles.com        & 2024-42  \\
    mmamania.com               & 2024-42  & realself.com              & 2024-42
    \\
    \bottomrule
    \end{tabular}}
  \caption{Domains that explicitly and fully allow GPTBot in their robots.txt, and the Common Crawl snapshot in which we first observed this behavior.}
    \label{tab:detail-allow}
  \end{table}

\section{Active Blocking}

\subsection{Squarespace Restricted AI Bots}
\label{appendix:squarespacebots}

The following directives are added to the robots.txt file for a
Squarespace site when a customer turns off the ``Artificial
Intelligence Crawlers'' option:

\begin{quote}
  \begin{verbatim}
    User-agent: GPTBot
    User-agent: ChatGPT-User
    User-agent: CCBot
    User-agent: anthropic-ai
    User-agent: Google-Extended
    User-agent: FacebookBot
    User-agent: Claude-Web
    User-agent: cohere-ai
    User-agent: PerplexityBot
    User-agent: Applebot-Extended
    Disallow: /
    \end{verbatim}
\end{quote}    

\subsection{Cloudflare ``Definitely Automated''}
\label{appendix:cloudflaredefautomated}

The following list shows the user agents we inferred Cloudflare's
``Definitely Automated'' setting to block:

\begin{quote}\begin{verbatim}
    360Spider           libwww-perl
    AHC                 magpie-crawler
    aiohttp             MeltwaterNews
    anthropic-ai        node-fetch
    Apache-HttpClient   Nutch
    axios               omgili
    binlar              PerplexityBot
    Bytespider          PhantomJS
    CCBot               PHP-Curl-Class
    centurybot          PiplBot
    Claudebot           python-requests
    curl                Python-urllib
    Diffbot             Scrapy
    Go-http-client      serpstatbot
    grub.org            Teoma
    HeadlessChrome      W3C-checklink
    httpx               wget
\end{verbatim}\end{quote}

We note that IP address likely plays a role in the operation of this
setting to block ``fake'' verified bots (e.g., a request that claims
to be a particular Cloudflare Verified Bot, but does not come from a
documented IP address).  We exclude these user agents from the list,
but note that the list of Cloudflare verified bots is publicly
available~\cite{cloudflareverifiedbots}.



\subsection{Cloudflare's ``Block AI Scrapers and Crawlers''}
\label{cloudflareblockaiscraperslist}

The following user agents are blocked by Cloudflare's ``Block AI Scrapers and Crawlers'' option:

\begin{quote}\begin{verbatim}
      Amazonbot            Diffbot/
      AwarioRssBot         GPTBot/
      AwarioSmartBot       magpie-crawler
      Bytespider           MeltwaterNews
      CCBot/               omgili/
      ChatGPT-User         PerplexityBot
      Claude-Web           PiplBot
      ClaudeBot            YouBot
      cohere-ai
\end{verbatim}\end{quote}
  
\noindent Note that AwarioRssBot, AwarioSmartBot, magpie-crawler, and
MeltwaterNews are not in the Dark Visitors list of AI user agents.

\section{Artist Survey}

\subsection{Survey Questions}
\label{appendix:survey}

In this section, we provide the list of questions that we asked in the artist survey. We omit the questions related to contact information and compensation. Our study was approved by our university's Institutional Review Board (IRB).
\\
\\
\small{
\textit{Questions about artistic background}

\textbf{Q1. Do you consider yourself a professional artist?}\\
\hspace*{3em}
\begin{itemize*}
    \item Yes
    \item No
\end{itemize*}

\textbf{Q2. What portion of your income comes from your art?}\\
\hspace*{3em}
\begin{itemize*}
    \item I haven't made any money from my art\\\hspace*{3em}
    \item I make some income from my art but it's not the main source\\\hspace*{3em}
    \item My art is my main source of income
\end{itemize*}

\textbf{Q3. How long have you been making money from your art?}\\
\hspace*{3em}
\begin{itemize*}
    \item Less than 1 year
    \item 1--5 years
    \item 5--10 years
    \item 10 years or more
\end{itemize*}

\textbf{Q4. What type of art do you do? (Select all that apply)}\\
\hspace*{3em}
\begin{itemize*}
    \item Concept Art
    \item Traditional Painting and Drawing
    \item Photography\\\hspace*{3em}
    \item Abstract
    \item Illustration
    \item Game Art
    \item Anime and Manga Art\\\hspace*{3em}
    \item Digital 2D
    \item Digital 3D
    \item Traditional Sculpting
    \item Environmental\\\hspace*{3em}
    \item Character and Creature Design
    \item Comicbook Art
    \item Matte Painting\\\hspace*{3em}
    \item Items Props
    \item Other (please specify)
\end{itemize*}

\textbf{Q5. Which country do you live in?}\\
\hspace*{3em}
\begin{itemize*}
    \item Australia
    \item Brazil
    \item Canada
    \item China
    \item France
    \item Germany
    \item India\\\hspace*{3em}
    \item Italy
    \item Japan
    \item Mexico
    \item Russia
    \item South Africa
    \item Spain\\\hspace*{3em}
    \item United Kingdom
    \item United States
    \item Other (please specify)
\end{itemize*}
\\
\\
\textit{Questions about technical background}

\textbf{Q6. How familiar are you with the following computer and internet items?} \emph{(1-5; 1 = no understanding, 5 = full understanding.)}\\\hspace*{3em}
\begin{itemize*}
    \item Website
    \item Generative AI
    \item Search engine\\\hspace*{3em}
    \item Nearest diffusion tree
    \item Robots.txt
\end{itemize*}

\textbf{Q7. Do you post your art online?}\\
\hspace*{3em}
\begin{itemize*}
    \item Yes
    \item No
\end{itemize*}

\textbf{Q8. Where do you post art online? (Select all that apply)}\\
\hspace*{3em}
\begin{itemize*}
    \item Social Media (Instagram, LinkedIn, \dots)\\\hspace*{3em}
    \item Art Platforms (ArtStation, DeviantArt, \dots)\\\hspace*{3em}
    \item Personal Website\\\hspace*{3em}
    \item Art Seller Websites (Artsy, Artrepreneur, \dots)\\\hspace*{3em}
    \item Other (please specify)
\end{itemize*}

\textbf{Q9. How do you host your personal website?}\\
\hspace*{3em}
\begin{itemize*}
    \item I have my own server
    \item Free service (e.g., free server with AWS)\\\hspace*{3em}
    \item Paid service (e.g., Squarespace with a custom domain)\\\hspace*{3em}
    \item Other (please specify)
\end{itemize*}

\textbf{Q10. What is the name of the service you use?}\\\hspace*{3em}
Answer: \rule{5cm}{0.15mm}

\textbf{Q11. Why did you choose the service?}\\\hspace*{3em}
Answer: \rule{5cm}{0.15mm}

\textbf{Q12. [Optional] If you're comfortable, please share a link to your personal website.}\\\hspace*{3em}
Answer: \rule{5cm}{0.15mm}
\\
\\
\textit{Questions about impressions of AI art and their actions}

\textbf{Q13. How familiar are you with AI-generated art?}\\
\hspace*{3em}
\begin{itemize*}
    \item Not familiar at all\\\hspace*{3em}
    \item Slightly familiar\\\hspace*{3em}
    \item Somewhat familiar\\\hspace*{3em}
    \item Moderately familiar\\\hspace*{3em}
    \item Very familiar
\end{itemize*}

\textbf{Q14. Do you use AI in your artistic process?}\\
\hspace*{3em}
\begin{itemize*}
    \item Never
    \item Rarely
    \item Sometimes
    \item Often
    \item Always
\end{itemize*}

\textbf{Q15. Please briefly describe your general impression of AI-generated art.}\\
\hspace*{3em}
Answer: \rule{5cm}{0.15mm}

\textbf{Q16. How much impact do you expect AI-generated art to have on your job security?}\\
\hspace*{3em}
\begin{itemize*}
    \item No impact
    \item Minor impact
    \item Moderate impact\\\hspace*{3em}
    \item Significant impact
    \item Severe impact
\end{itemize*}

\textbf{Q17. Have you taken any actions because of the increasing use of AI-generated art in recent years?}\\
\hspace*{3em}
\begin{itemize*}
    \item Yes
    \item No
\end{itemize*}

\textbf{Q18. What actions have you taken? (Select all that apply)}\\
\hspace*{3em}
\begin{itemize*}
    \item Reducing the amount of my artwork that I share online\\\hspace*{3em}
    \item Actively removing my old artwork from the Internet\\\hspace*{3em}
    \item Posting lower resolution versions of my artwork online\\\hspace*{3em}
    \item Learning about AI art tools and possibly using them\\\hspace*{3em}
    \item Preventing my websites from being scraped\\\hspace*{3em}
    \item Using Glaze to protect my art before posting\\\hspace*{3em}
    \item Other (please specify)
\end{itemize*}

\textbf{Q19. Please elaborate on how you prevent your websites from being scraped.}\\
\hspace*{3em}
Answer: \rule{5cm}{0.15mm}

\textbf{Q20. Do you plan to take any actions because of the increasing use of AI-generated art in recent years?}\\
\hspace*{3em}
\begin{itemize*}
    \item Yes
    \item No
\end{itemize*}

\textbf{Q21. What actions do you plan to take? (Select all that apply)}\\
\hspace*{3em}
\begin{itemize*}
    \item Reducing the amount of my artwork that I share online\\\hspace*{3em}
    \item Actively removing my old artwork from the Internet\\\hspace*{3em}
    \item Posting lower resolution versions of my artwork online\\\hspace*{3em}
    \item Learning about AI art tools and possibly using them\\\hspace*{3em}
    \item Using Glaze to protect my art before posting\\\hspace*{3em}
    \item Preventing my websites from being scraped\\\hspace*{3em}
    \item Other (please specify)
\end{itemize*}

\textbf{Q22. If your website hosting platform offers a mechanism (e.g. by clicking a button) to tell AI companies that you would like them not to scrape your website, how likely will you enable this mechanism?}\\
\hspace*{3em}
\begin{itemize*}
    \item Not likely at all
    \item Unlikely
    \item Neutral / Undecided \\\hspace*{3em}
    \item Likely
    \item Very likely
\end{itemize*}\\
\hspace*{1.5em}\textbf{Why or why not?} Answer: \rule{3cm}{0.15mm}

\textbf{Q23. If your website hosting platform offers a mechanism (e.g. by clicking a button) to block AI companies from scraping your website, how likely will you enable this mechanism?}\\
\hspace*{3em}
\begin{itemize*}
    \item Not likely at all
    \item Unlikely
    \item Neutral / Undecided\\\hspace*{3em}
    \item Likely
    \item Very likely
\end{itemize*}\\
\hspace*{1.5em}\textbf{Why or why not?} Answer: \rule{3cm}{0.15mm}
\\
\\
\textit{Questions about knowledge of robots.txt}

\textbf{Q24. Have you heard about robots.txt before today?}\\
\hspace*{3em}
\begin{itemize*}
    \item Yes
    \item No
\end{itemize*}

\textbf{Description of robots.txt for artists who select ``no'' in Q24. This description is generated with the help of ChatGPT.}\\
\hspace*{3em}
Do you know that over 90\% of artists don't realize they can use a simple tool called robots.txt to stop automated programs (also known as bots) from downloading content from their websites? Think of robots.txt as a ``Do Not Enter'' sign for automated programs that browse the internet. When placed on a website, it tells these automated programs which parts of the site they're not allowed to access. While it won't stop every bot, it works like a polite request to keep things like personal galleries or portfolios hidden from search engines or unwanted bots. This is an easy way for artists to protect their work and control how it appears online, without needing to dive into complicated tech or legal steps. Adding a robots.txt file can be a quick win for maintaining privacy and keeping unwanted eyes off your art.\\\hspace*{3em}
That being said, it is important to note that not all companies respect robots.txt---some may ignore it entirely if they choose to.

\textbf{Q25. Briefly describe what you think robots.txt does.}\\\hspace*{3em}
Answer: \rule{5cm}{0.15mm}

\textbf{Q26. Would you consider adopting robots.txt in the future?}\\
\hspace*{3em}
\begin{itemize*}
    \item Not likely at all
    \item Unlikely
    \item Neutral / Undecided\\\hspace*{3em}
    \item Likely
    \item Very likely
\end{itemize*}\\\hspace*{1.5em}
\textbf{Why or why not?} \emph{(Open-ended)}

\textbf{Q27. Robots.txt is a standardized way to declare ``do not crawl,'' and most companies respect it. How likely do you think AI companies will respect robots.txt?}\\
\hspace*{3em}
\begin{itemize*}
    \item Not likely at all
    \item Unlikely
    \item Neutral / Undecided\\\hspace*{3em}
    \item Likely
    \item Very likely
\end{itemize*}\\\hspace*{1.5em}
\textbf{Why or why not?} Answer: \rule{3cm}{0.15mm}

\textbf{Q28. Have you checked the robots.txt of websites where you post your work?}\\
\hspace*{3em}
\begin{itemize*}
    \item Yes
    \item No
\end{itemize*}

\textbf{Q29. Can you control (edit or modify) the content of the robots.txt of websites where you post your work?}\\
\hspace*{3em}
\begin{itemize*}
    \item I have full control over the full content of robots.txt\\\hspace*{3em}
    \item I can click some buttons to switch between a few presets\\\hspace*{3em}
    \item I have no control over the content\\\hspace*{3em}
    \item I am not sure\\\hspace*{3em}
    \item Other (please specify)
\end{itemize*}

\textbf{Q30. How did you get the current content of robots.txt?}\\
\hspace*{3em}
\begin{itemize*}
    \item Provided by my website hosting platform\\\hspace*{3em}
    \item Copied from the Internet (e.g., a blog)\\\hspace*{3em}
    \item Created my own robots.txt\\\hspace*{3em}
    \item Other (please specify)
\end{itemize*}

\textbf{Q31. Do you currently use robots.txt to disallow bots from AI companies from scraping websites where you post your art?}\\
\hspace*{3em}
\begin{itemize*}
    \item Yes
    \item No
\end{itemize*}\\\hspace*{1.5em}
\textbf{Why?} Answer: \rule{3cm}{0.15mm}\\\hspace*{1.5em}
\textbf{Why not?}\\\hspace*{3em}
\begin{itemize*}
    \item I am concerned it will impact the discoverability of my website online\\
    \hspace*{3em}
    \item I don't mind AI training on my art\\
    \hspace*{3em}
    \item I don't know how to do it\\
    \hspace*{3em}
    \item Other (please specify)
\end{itemize*}
\\\hspace*{1.5em}
\textbf{Q32. [Optional] Do you face any obstacles in adopting robots.txt? (Select all that apply)}\\
\hspace*{3em}
\begin{itemize*}
    \item I have trouble finding how to edit the robots.txt\\
    \hspace*{3em}
    \item I find it hard to write the robots.txt\\
    \hspace*{3em}
    \item I don't know how to use it\\
    \hspace*{3em}
    \item Other (please specify)
\end{itemize*}
\\
\\
}

\subsection{Demographics}
\label{appendix:demographics}
\normalsize

This section presents the demographics of the participants in our survey. As previously mentioned, we focus on their artistic background, as it is the most  relevant to our study. 

\begin{table}[ht]
\centering
\begin{tabular}{lr}
\toprule
\textbf{Duration} & \textbf{Count} \\
\midrule
Less than 1 year & 17  \\
1--5 years       & 68  \\
5--10 years      & 44  \\
10 years or more & 47  \\
\midrule
\textbf{Total}   & \textbf{176} \\
\end{tabular}
\caption{How long participants have been making money from their art.\label{tab:making_money_years}}
\end{table}

Table~\ref{tab:making_money_years} presents a breakdown of how long participants have been making money from their art.  The majority of respondents (68) have been doing so for 1--5 years, whereas only 17 have been making money from their art for less than a year. Over half of the respondents (91) have been making money from their art for at least 5 years.


\begin{table}[ht]
\centering
\begin{tabular}{lr}
\toprule  
\textbf{Continent} & \textbf{Count} \\
\midrule
North America & 109 \\
Europe        & 52  \\
Asia          & 21  \\
South America & 18  \\
Africa        & 2   \\
Oceania       & 1   \\
\midrule
\textbf{Total}   & \textbf{203} \\
\end{tabular}
\caption{Continent of residence of participants.\label{tab:continent}}
\end{table}
Table~\ref{tab:continent} presents a breakdown of the continent of residence of participants. The majority of participants (109) are from North America, with 89 of them from the United States. The second largest group is from Europe (52), with 18 from the United Kingdom, five from Poland, and another five from Germany. The third largest group is from Asia (21), with nine from The Philippines. The remaining participants are from South America~(18), Africa~(2), and Oceania~(1).

\begin{table}[ht]
\centering
\begin{tabular}{lr}
\toprule  
\textbf{Art Type} & \textbf{Count} \\
\midrule
Illustration                         & 163 \\
Digital 2D                           & 143 \\
Character and Creature Design        & 99  \\
Traditional Painting and Drawing     & 78  \\
Concept Art                          & 68  \\
\midrule
\textbf{Total}   & \textbf{551} \\
\end{tabular}
\caption{Top five types of art participants do.\label{tab:art_type}}
\end{table}

Table~\ref{tab:art_type} presents a breakdown of the top five types of art participants do. Each participant can select every type of art they do, so the total number of responses is greater than the number of participants. The most common type of art is illustration (163), followed by digital 2D (143), character and creature design (99), traditional painting and drawing (78), and concept art (68).

\begin{table}[ht]
\centering
\begin{tabular}{lr}
\toprule  
\textbf{Term} & \textbf{Average Familiarity} \\
\midrule
Website                 & 4.60 \\
Search Engine           & 4.35 \\
Generative AI           & 3.89 \\
Robots.txt              & 1.99 \\
\textit{Nearest diffusion tree}  & 1.56 \\
\midrule
\end{tabular}
\caption{Participant's average familiarity with various terms. The average is on a scale from 1 to 5, where 1 represents no understanding and 5 represents full understanding. Following the work of Hargittai~\cite{hargittai2009update}, we also include a bogus item ``Nearest diffusion tree'', indicated in italics.
\label{tab:familiarity}}
\end{table}

Table~\ref{tab:familiarity} presents our participant's average familiarity with various terms. This question is designed to assess our participants's digital literacy. The average is on a scale from 1 to 5, where 1 represents no understanding and 5 represents full understanding. Following the work of Hargittai~\cite{hargittai2009update}, we also include a bogus item ``Nearest diffusion tree'', indicated in italics. 
 The most familiar term is ``website'' (4.60), followed by ``search engine'' (4.35), ``generative AI'' (3.89), and ``robots.txt'' (1.99). The least familiar term is ``nearest diffusion tree'' (1.56). Given that this bogus term was rated as the lowest compared to the other four terms, we conclude that our participants do not select randomly. This data also suggests that our participants are relatively familiar with general terms such as ``website'', ``search engine'', and ``generative AI'', but much less familiar with ``robots.txt''. This result is consistent with our other findings in Section~\ref{sec:artist-sentiment}.

\newpage
\onecolumn
\subsection{Codebook}
\label{appendix:codebook}
This section details the codebook we used to analyze the qualitative data collected from artists. Specifically, Table~\ref{tab:other-actions} lists other actions taken by artists in response to AI-generated art; Table~\ref{tab:robots-not} lists reasons why artists would not adopt robots.txt; Table~\ref{tab:yes-block} lists reasons why artists would enable a mechanism that blocks AI crawlers; and Table~\ref{tab:trust} lists reasons why artists do not trust AI companies to respect robots.txt.

\begin{table*}[!htbp]
\centering
\begin{tabular}{p{0.15\linewidth} p{0.27\linewidth} p{0.48\linewidth}}
\toprule
\textbf{Theme} & \textbf{Description} & \textbf{Example} \\ \midrule
Modify post & Artists alter the content or format of the artwork they share online. & ``Overlaying watermarks or art filters to modify the artwork'' \\[0.3em]

Switch platforms & Artists migrate to alternative sites or remove their work from certain platforms. & ``Use Cara instead of Instagram'' \\[0.3em]

Raise awareness & Artists publicly highlight issues affecting them or the community. & ``Spreading awareness about the damage AI-generated art does'' \\[0.3em]

Unionize & Artists organize collectively to negotiate or advocate for shared interests. & ``Connecting with groups of professional artists being impacted to search for collective solutions for our field'' \\[0.3em]

Change career path & Artists pivot to a different professional direction. & ``I left school and am taking a gap year to reevaluate my life'' \\[0.3em]

Miscellaneous & Additional strategies not covered above. & ``Using block lists to block AI art accounts'' \\ \bottomrule
\end{tabular}
\caption{Codebook for other actions taken by artists in response to AI-generated art.\label{tab:other-actions}}
\end{table*}

\begin{table*}[!htbp]
\centering
\begin{tabular}{p{0.15\linewidth} p{0.27\linewidth} p{0.48\linewidth}}
    \toprule
\textbf{Theme} & \textbf{Description} & \textbf{Example} \\ \midrule
Efficacy & Artists are concerned about the efficacy of robots.txt given its voluntary nature. & ``if the companies can ignore it why would they respect it considering what they already do'' \\[0.3em]

Usability & Artists are concerned about the complexity of implementing or using robots.txt. & ``It sounds like something difficult to use'' \\[0.3em]

More information & Artists want to gather more information about robots.txt before making a decision. & ``Not informed enough about it'' \\[0.3em]

No personal website & Artists do not have a personal website. & ``I do not have a personal website'' \\[0.3em]

Search results & Artists are concerned about robots.txt impacting the search results of their websites. & ``If it hides things from *search engines* then how will people find my work?...'' \\[0.3em]

\bottomrule
\end{tabular}
\caption{Codebook for why artists would not adopt robots.txt.\label{tab:robots-not}}
\end{table*}

\begin{table*}[h]
\centering
\begin{tabular}{p{0.15\linewidth} p{0.27\linewidth} p{0.48\linewidth}}
\toprule
\textbf{Theme} & \textbf{Description} & \textbf{Example} \\ \midrule
Protection & Artists want to protect their work. & ``To protect my original concepts and visual brand (aka original character designs and artstyle)''\\[0.3em]

Consent & Artists do not want their work to be crawled and do not consent to crawling. & ``I havent given AI companies permission to use my work'' \\[0.3em]

Compensation & Artists are not compensated while AI companies profit from their work. & ``..., and I do not want other companies to profit off of it without my knowledge, permission, or without fair compensation towards the source.'' \\[0.3em]

Useful mechanism & Artists see this mechanism as useful and reassuring. & ``Adds a sense of security and ease of use.'' \\[0.3em]

Legal benefit & Artists believe such mechanisms could be potentially useful in legal cases. & ``..., it is a measure to reinforce a statement that we do not condone with these practices and will probably benefit in a possible lawsuit in the future.'' \\[0.3em]

Misc & Additional reasons not covered above. & ``At this point if the option is presented I'll do my research on it and if it seems legitimate I'll do it on principle.'' \\[0.3em]

\bottomrule
\end{tabular}
\caption{Codebook for why artists would enable a mechanism that blocks AI crawlers.\label{tab:yes-block}}
\end{table*}
    
\begin{table*}[!htbp]
\centering
\begin{tabular}{p{0.15\linewidth} p{0.27\linewidth} p{0.48\linewidth}}
    \toprule
\textbf{Theme} & \textbf{Description} & \textbf{Example} \\ \midrule
Track record & AI companies have a history of conducting operations that maybe unauthorized and unethical. & ``Based on the attitudes I have seen from AI companies and the way AI companies have already used data without consent, I'm unsure if they will respect robot.txt'' \\[0.3em]

Profit & AI companies have monetary interests in scraping artists' work. & ``Money before morals.'' \\[0.3em]

Perception & Artists perceive AI companies negatively (e.g., as greedy or unethical). & ``AI companies are morally bankrupt.''  \\[0.3em]

Loophole & AI companies might find loopholes or workarounds to bypass robots.txt.& ``They might start loopholes to get around it or something '' \\[0.3em]

Legal enforcement & The need and lack of legislation or legal enforcement. & ``Generative AI is built on top of copyright infringement-they can't be profitable without it, so they will argue against any thing that prevents them from scrapping. They have to be forced to respect it by law, we can't trust their good faith.'' \\[0.3em]

Voluntary nature & Robots.txt is a voluntary mechanism. & ``At best it seems that robot.txt is just a warning sign, and will not entirely stop AI companies from deciding to scrape any particular content.'' \\[0.3em]

Misc & Additional reasons not covered above. & ``I think, unfortunately, a lot of companies will not respect and will do it anyway.'' \\[0.3em]

\bottomrule
\end{tabular}
\caption{Codebook for why artists do not trust AI companies to respect robots.txt.\label{tab:trust}}
\end{table*}

\end{document}